\documentstyle[12pt]{article}

\font\twelvemsb=msbm10 scaled 1200 
\def\Bbb#1{\hbox {\twelvemsb#1}}

\newcommand{\M}{{\cal M}}
\newcommand{\Minf}{{\cal M}_{\infty}}
\newcommand{\haMinf}{\hat{\cal M}_{\infty}}
\newcommand{\Mun}{{\cal M}_f}
\newcommand{\gun}{g_f}
\newcommand{\MI}{\M_2}

\newcommand{\MN}{\stackrel{o}{\M_{1}}}

\newcommand{\MD}{\M_3^D}
\newcommand{\MPz}{\M_3^0}
\newcommand{\MPq}{\M_3^4}
\newcommand{\Mk}{\M_3^{k}}
\newcommand{\Ml}{\M_3^{l}}
\newcommand{\Mpi}{\M_3^{\pi}}

\newcommand\bm[1]{\mbox{\boldmath$#1$}}
\newcommand\cm[1]{\bm{\cal#1}} 
\newcommand{\C}{\cm{C}} 
 
\newcommand\ov[1]{\overline{#1}} 
\newcommand{\Fsq}{\bm{{\cal F}^2}}
\newcommand{\F}{\bm{{\cal F}}}  
\newcommand{\FF}{\bm{F}}

\newcommand{\I}{{\cm I}}

\newtheorem{lemma}{Lemma}
\newtheorem{theorem}{Theorem}
\newtheorem{definition}{Definition}
\newtheorem{proposition}{Proposition}
\begin{document}

\title{Uniqueness properties of the Kerr metric}
\author{Marc Mars\thanks{Also at Laboratori de F\'{\i}sica Matem\`atica,
Societat Catalana de F\'{\i}sica, IEC, Barcelona,Spain.}
\\  Departament de
F\'{\i}sica Fonamental, Universitat de Barcelona, \\
Av. Diagonal 647, 08028 Barcelona, Spain. \\
Fax. + 34 93 402 11 49, e-mail: marc@fismat.ffn.ub.es}
\maketitle
\begin{abstract}
We obtain a geometrical condition on vacuum, stationary,
asymptotically flat
spacetimes which is necessary and sufficient for
the spacetime to be locally isometric to Kerr.
Namely, we prove
a theorem stating that an asymptotically flat, stationary, vacuum
spacetime  such that the so-called Killing form is an eigenvector
of the self-dual Weyl tensor must be locally isometric to Kerr.
Asymptotic flatness
is a fundamental hypothesis of the theorem, as we demonstrate by
writing down the family of metrics obtained when this requirement
is dropped.
This result indicates why the
Kerr metric plays such an important role  in general relativity. It may
also be of interest in order to extend the uniqueness theorems of black
holes to the non-connected and to the non-analytic case.

\end{abstract}


\newpage

\section{Introduction}

The Kerr metric is of fundamental importance in
general relativity. Its relevance comes mainly
from the uniqueness theorem for black
holes, which states that, under rather general conditions,
the Kerr spacetime is the only asymptotically flat, stationary,
vacuum black hole. Therefore, the Kerr metric  describes the
exterior endstate  of any sufficiently massive
collapsing isolated system which obeys the cosmic censorship conjecture
(assuming that some equilibrium state is reached).
Despite its importance, a clear understanding of why the Kerr metric is
so special remains incomplete.
In comparison, the Schwarzschild metric has nice
uniqueness properties like Birkhoff's theorem or the conformal
flatness of the hypersurfaces of constant static time (this second property
characterizes Schwarzschild among static, asymptotically flat, vacuum
spacetimes).

Obtaining a similar characterization for the Kerr metric has interest
not only in order to understand better 
the Kerr metric
but also in order to try and generalize
the black hole uniqueness theorem. Indeed, the known proofs for
this theorem (in the non-static case) require hypotheses
which are too strong both from the mathematical and from the physical
point of view. First of all, connectedness of the
event horizon (i.e. the existence of a single black hole)
is a fundamental hypothesis.	
Dropping this condition is necessary in order
to exclude the existence of an equilibrium configuration
containing several black holes.
In the static case, the non-connected case
was solved  by G.\ L.\ Bunting and  Masood-ul-Alam \cite{BM} by exploiting
the conformal flatness of the static slices
of the Schwarzschild spacetime.
Another requirement
in the existing proofs is the analyticity of the metric and of the event
horizon (see Chru\'sciel  \cite{Ch3}), which is
clearly a very strong and little justified assumption.  
Analyticity is used in order to apply  the Hawking rigidity
theorem \cite{HE} which
proves the existence of a second Killing vector. Since this is an
early step in the proof, relaxing the
analyticity condition would
probably force a completely new approach to the problem.
Although the black hole uniqueness
theorem is generally
believed to be true when these two conditions are relaxed,
few results in this direction are known at present
(see, however,  Weinstein
\cite{We1}, \cite{We2} for some progress in the non-connected case).
We believe that
obtaining a suitable characterization of the Kerr metric among stationary,
asymptotically flat, vacuum metrics would be an important step forward in this
problem.

Recently, using a  characterization of the Kerr metric in terms of
the so-called Simon tensor \cite{Simon} in the manifold
of trajectories, we have been able \cite{Mars1} to find the following
geometric characterization of the Kerr metric.
Let us consider the class of vacuum spacetimes
admitting a Killing vector (with no restriction on its causal character).
From the Killing vector
$\vec{\xi}$, we construct the so-called Killing form $\F_{\alpha\beta}
\equiv \nabla_{\alpha} \xi_{\beta} + \frac{i}{2} \eta_{\alpha\beta\lambda\mu}
\nabla^{\lambda} \xi^{\mu}$ ($\eta_{\alpha\beta\lambda\mu}$ is the
volume form) and the
Ernst potential $\chi \equiv \lambda - i \omega$ where
$\lambda = - \xi^{\alpha} \xi_{\alpha}$ and
$\nabla_{\alpha} \omega =
\eta_{\alpha\beta\mu\nu} \xi^{\beta} \nabla^{\mu} \xi^{\nu}$ ($\omega$
and hence $\chi$ are defined only locally). $\F_{\alpha\beta}$ is
a self-dual 2-form, i.e. it satisfies $\F^{\star} = - i \F$ where
$\star$ is the Hodge dual operator. On the other hand,
we consider the self-dual Weyl tensor from the right
$\C_{\alpha\beta\lambda\mu}
\equiv C_{\alpha\beta\lambda\mu} + \frac{i}{2} \eta_{\lambda\mu\rho\sigma}
C_{\alpha\beta}^{\,\,\,\,\,\,\,\rho\sigma}$
viewed as a symmetric tensor in the space
of self-dual two forms (as is usually done in obtaining the Petrov type
of a spacetime, see e.g. \cite{KSMH}).
Thus, we have at hand two geometrical
objects which are, in general, unrelated to each other. The
theorem in \cite{Mars1} states, roughly speaking, that
the Kerr metric is characterized by the appropriate ``matching'' of these
two objects.
More precisely,

\begin{theorem}
Let $(\M,g)$ be a smooth, vacuum spacetime admitting a Killing
vector $\vec{\xi}$. Let the Killing form be defined as above and let $\M$
satisfy
\begin{enumerate}
\item There exists a non-empty region $\Mun$ where $
\Fsq \equiv \F_{\alpha\beta}
\F^{\alpha\beta} \neq 0$.
\item The Killing form and the Weyl tensor are related by
\begin{eqnarray}
\C^{\alpha\beta}_{\,\,\,\,\,\,\,\,\gamma\delta}  =
H  \left (\F^{\alpha\beta} \F_{\gamma\delta} - \frac{1}{3}
\I^{\alpha\beta}_{\,\,\,\,\,\,\,\,\gamma\delta} \Fsq\right )
\label{defining}
\end{eqnarray}
where $\I^{\alpha\beta}_{\,\,\,\,\,\,\,\,\gamma\delta}$ is the identity
on the space of self-dual 2-forms and $H$ is a scalar function.
\end{enumerate}
Then, there exist two complex constants $l$ and $c$ such that
$H = - 6/(c - \chi)$ and
$\F^2 = - l (c - \chi )^4$.

If, in addition,  $c = 1$ and $l$ is
real and positive, then  $(\M,g)$ is locally isometric to the
Kerr spacetime.
\label{Old}
\end{theorem}

In reference \cite{Mars1}, this theorem was stated in a different form.
There, the hypotheses included asymptotic flatness of the spacetime
and the vanishing of the so-called spacetime Simon tensor. However, as
it was noticed at the end of Remark 1, asymptotic flatness was used only
in order to prove Lemma 5 in that paper, which just fixes the value
of two complex constants. On the other hand, the vanishing
of the spacetime Simon tensor can be replaced by hypothesis 2 above by
virtue of lemma 4 in \cite{Mars1}. It is easy to see that theorem
\ref{Old} holds following essentially the same steps as in the
proof of the theorem stated in \cite{Mars1}. We have preferred to state 
the theorem in this new form
in order to emphasize that asymptotic flatness plays  
no role in the  characterization of Kerr obtained in \cite{Mars1}
(i.e. the characterization is purely local). If asymptotic flatness
is imposed on $(\M,g)$ then condition 1 in the theorem can be dropped
and the constants $l$ and $c$ appearing in condition 2 are fixed
automatically to its values $c=1$ and $l>0$ by a suitable rescaling of
the Killing vector.

Theorem \ref{Old} characterizes Kerr in a neat and geometrical
way. However, it is ``too local'' for our purposes. 
Indeed, the condition of asymptotic flatness on the spacetime is
of fundamental importance in the proof of the black hole uniqueness theorem
(if this condition is dropped, then the uniqueness of the Kerr black hole
is not true). On the other hand, the characterization of Kerr in
theorem \ref{Old} does not use asymptotic flatness at all. 
This suggests very strongly
that asymptotic flatness has not been 
exploited in its full extent in order to characterize the Kerr metric
among stationary, vacuum  and {\it asymptotically flat} spacetimes.
It is plausible to believe that by
employing asymptotic flatness in a more essential way, the
geometric condition (\ref{defining})
involving the Weyl tensor and the Killing form
can be relaxed significantly.
In that case, we would certainly be in a better
position for generalizing the black hole uniqueness theorems.

In this paper we present a generalization of Theorem \ref{Old} along these
lines. The main theorem in this paper is described in the next
subsection, while its proof is left for the following sections.

\subsection{Main result and discussion}

The first task to consider is trying to guess which condition
should replace (\ref{defining}). To do that let us describe
in some more detail the geometric content of (\ref{defining}).
An arbitrary self-dual 2-form like $\F_{\alpha\beta}$ can be
algebraically classified by analyzing its associated eigenvalue problem.
At points where $\F_{\alpha\beta}$ is regular (i.e. $\Fsq \neq 0$),
there exist two principal
null directions defined as the real eigenvectors of $\F_{\alpha\beta}$, i.e.
\begin{eqnarray}
\F^{\alpha}_{\,\,\,\beta} l^{\beta} \propto l^{\alpha}.
\label{eigen}
\end{eqnarray}
At points
where $\F_{\alpha\beta}$ is singular (i.e $\Fsq =0$ with $\F_{\alpha\beta}
\neq 0$) there exists one principal null direction.
Similarly, the Petrov classification of the Weyl tensor
is the algebraic classification of the
endomorphism of the space of self-dual 2-forms defined by
$\C \left  ({\cm X} \right )_{\alpha\beta} =
\C_{\alpha\beta}^{\,\,\,\,\,\,\,\,\mu\nu} {\cm X}_{\mu\nu}$.
There exist four principal null directions
of the Weyl tensor (which  degenerate when the
Weyl tensor is algebraically special). A repeated
principal null direction of the Weyl tensor is a non-zero
null vector field $\vec{l}$ satisfying
$\C_{\alpha\beta\gamma\delta} l^{\beta} l^{\delta} \propto
l_{\alpha} l_{\gamma}$.
The Petrov classification of
a Weyl tensor satisfying (\ref{defining}) is very simple.
The Petrov type is $D$ whenever $H \Fsq  \neq 0$,
$N$ when $H \neq 0$, $\Fsq =0$ and $0$ when $H=0$. Moreover, at points
where $\F_{\alpha\beta}$
is regular, each principal null direction of $\F_{\alpha\beta}$ is a
double principal null direction of the Weyl tensor (this is an
alternative way of writing down the characterization of Kerr given
in Theorem \ref{Old}).
So, Theorem \ref{Old} involves a condition on the Weyl tensor that
already determines its Petrov type. This is a strong local condition.
We would like to generalize it so that the Petrov type remains 
undetermined. A possible way of doing this is by demanding
\begin{eqnarray}
\C^{\alpha\beta}_{\,\,\,\,\,\,\,\,\gamma\delta} 
\F^{\gamma\delta} \propto \F^{\alpha\beta}.
\label{defining2}
\end{eqnarray}
This condition is obviously satisfied by (\ref{defining}) but it
is much weaker than (\ref{defining}). Indeed, while
condition (\ref{defining}) restricts the full form of the Weyl
tensor in the spacetime (and therefore imposes strong conditions on the
Petrov type), condition (\ref{defining2}) demands just that
the Killing form is an eigenvector of the self-dual Weyl tensor. A priori,
this condition does not restrict the Petrov type of the spacetime
whatsoever. Moreover, the essence of condition (\ref{defining})
(i.e. that the two geometrical objects
$\F_{\alpha\beta}$ and $\C_{\alpha\beta\gamma\delta}$ ``match'' together)
is retained. This indicates that (\ref{defining2}) may be the
generalization we are seeking. The main objective
of this paper is proving the validity of this result.
The precise theorem we prove is

\begin{theorem}
Let $(\M,g)$ be a smooth, vacuum  spacetime with the following
properties
\begin{enumerate}
\item $(\M,g)$ admits a Killing field $\vec{\xi}$ such that its Killing
form $\F_{\alpha\beta}$ is an eigenvector of the self-dual
Weyl tensor $\C_{\alpha\beta\gamma\delta}$
\begin{eqnarray}
\C^{\alpha\beta}_{\,\,\,\,\,\,\,\,\gamma\delta} 
\F^{\gamma\delta} \propto \F^{\alpha\beta}.
\label{defining3}
\end{eqnarray}
\item $(\M,g)$ contains a stationary, asymptotically flat four-end
$\Minf$, $\vec{\xi}$ tends to a time translation at infinity
in $\Minf$ and the Komar mass of $\vec{\xi}$ in $\Minf$
is non-zero.
\end{enumerate}
Then $(\M,g)$ is locally isometric to a Kerr spacetime.
\label{Main}
\end{theorem}

{\bf Remark 1}. A stationary asymptotically flat four-end is 
an open submanifold $\Minf \subset \M$ diffeomorphic to
$I\times \left (\Bbb{R}^3 \setminus B(R) \right )$, ($I \in \Bbb{R}$
is an open interval and $B(R)$ is a closed ball of radius $R$), such that,
in the local coordinates $\{t,x^i \}$ 
defined by the diffeomorphism, the metric
satisfies
\begin{eqnarray*}
\left |g_{\mu\nu} - \eta_{\mu\nu} \right | +
\left |r \partial_i g_{\mu\nu} \right | \leq C r^{-\alpha},
\hspace{1cm} \partial_t g_{\mu\nu} =0
\end{eqnarray*}
where $C,\alpha$ are positive constants, $r = \sqrt{ \sum (x^i)^2}$ and
$\eta_{\mu\nu}$ is the Minkowski metric.  
Usually, the definition of asymptotically flat four-end requires 
$I = \Bbb{R}$ but we do not need this restriction.
The Einstein field equations together with the
existence of a timelike Killing vector force $\alpha \geq 1$
(Kennefick and \'{O} Murchadha \cite{KO}). Then, it
is well-known (see e.g. \cite{BS1}) that
the metric in the asymptotic region can be written in the form
\begin{eqnarray}
g_{00} = -1 + \frac{2M}{r} + O(r^{-2}),
\hspace{6mm}
g_{0i} = - \epsilon_{ijk} \frac{4 S^j x^k}{r^3} + O(r^{-3}),
\hspace{6mm} g_{ij} = \delta_{ij} + O(r^{-1}), 
\label{AF}
\end{eqnarray}
where $M$ is the Komar mass \cite{Ko} 
of $\vec{\xi}$ in the asymptotically flat end $\Minf$ (and hence
non-zero by assumption) and $\epsilon_{ijl}$ is the alternating Levi-Civita
symbol.

{\bf Remark 2}. The Kerr spacetime is understood to be
the maximal analytic
extension of the Kerr metric, as described by
Boyer and Lindquist \cite{BY} and Carter \cite{Carter2}.
An element of the Kerr family will be denoted by $(\M_{M,a},g_{M,a})$,
where $M$ denotes the Komar mass and
$a$ the specific angular momentum. The particular case when $a=0$
corresponds to the Kruskal extension of the Schwarzschild spacetime.

{\bf Remark 3}. 
The conclusion of the theorem is that $(\M,g)$ is locally isometric
to the Kerr spacetime. This concept is standard;
it means that for any point $p \in \M$, there
exists an open neighbourhood $U_p$
of $p$ which is isometrically diffeomorphic to an open submanifold
of $\M_{M,a}$.
Despite the fact that the characterization of Kerr in the
theorem involves a local condition (\ref{defining3})
and a global condition (asymptotic flatness),
we should not expect in principle 
that the local isometry
extends to an isometric embedding of $(\M,g)$ into $(\M_{M,a},g_{M,a})$.
The reason is that there may exist suitable identifications 
in the Kerr spacetime that define a spacetime which is still
asymptotically flat in the sense we are using (and the local
condition $1$ would obviously be still satisfied). 
Let us emphasize, however,
that Theorem \ref{Main} is ``semi-local'' 
because the existence of the local isometry is shown {\it everywhere}.
Actually, showing that the result holds
everywhere constitutes a substantial part of the proof.

While the methods in the proof of Theorem \ref{Main} are similar
to those in Theorem \ref{Old} the complexity of the proof increases
notably. The reason is, of course, that the local assumptions are
now much weaker and instead 
asymptotic flatness must be exploited in a stronger way.
In particular, the local
condition (\ref{defining3}) does not restrict the Petrov type and
several cases must be analyzed. Moreover, the interplay between
different regions where the Petrov type may change requires
special care. We refer
to Hall \cite{Hall} (see also \cite{Raul} for a generalization to other
continuous endomorphisms) for general restrictions on the regions
with different Petrov types. 

Our main objective in this paper is obtaining a characterization of
the Kerr metric among stationary, vacuum, asymptotically flat spacetimes
which uses the asymptotic properties in a fundamental way. To show that
Theorem \ref{Main} achieves this, we will prove the theorem
trying to separate the use of the local conditions from the use of
asymptotic flatness. This will be particularly so in Proposition \ref{funda}
below where the local condition 1 is imposed but 
asymptotic flatness is not used at all.
This proposition gives, essentially, the local form of the metric
on the region where the Petrov type is II (or D or 0).
This family of metrics is rather large and contains arbitrary 
functions in general. This shows that the local condition 1 is weak (it is
fulfilled by a large family of metrics) and only when combined with
asymptotic flatness gives a characterization of Kerr.
We consider it difficult to relax further
the local condition 1 and still maintain the conclusion of
the theorem. So, in some sense, asymptotic flatness has been exploited
``as much as possible'' to characterize Kerr. An eventual proof
of the black hole uniqueness theorem would require exploiting
the existence of a regular black hole in order to try and prove
that the local condition 1 holds. This problem is under current
investigation.

The paper is organized as follows. In section 2, we introduce our
notation and write down several identities which are
valid on any vacuum spacetime admitting a
Killing vector $\vec{\xi}$. These identities and definitions
will be used thoroughly in this paper.
We also write down some of the
consequences of (\ref{defining3}) and discuss the different
possibilities for the Petrov type at points where $\Fsq \neq0$.

The strategy of the proof consists in defining first an open, connected
and asymptotically flat subset $\Mun \subset \M$ where $\Fsq \neq 0$
everywhere and restricting the analysis to $\left ( \Mun, g |_{\Mun}
\right )$. The bulk of the proof is showing that the conclusion of
the theorem holds on $\Mun$. Afterwards, we prove that $\Mun = \M$.
In section 3, we show that the region
$\Mun$ must be of Petrov type II, D or 0 (thus excluding Petrov types III, N
and I on $\Mun$). Actually, Petrov type III is easily seen to be
incompatible with (\ref{defining3}) on $\Mun$, so only Petrov types N
and I must be studied. Petrov type N is dealt with easily 
but excluding Petrov type I requires a rather involved argument
that already uses several
properties of asymptotic flatness. In section 4 we analyze in detail the
Petrov type II (and its particularization to Petrov type D). First
we find the local form of the metric when the condition of asymptotic flatness
is dropped. As discussed above, the resulting family of metrics is rather
large . The remaining part of section 4 is devoted to show
that  $\left ( \Mun, g |_{\Mun} \right )$ is locally isometric to Kerr.
Finally we prove that $\Mun = \M$.

Before starting with the details, let us make two final
comments.
First, the proof of the theorem uses, obviously, the Einstein vacuum 
equations extensively.
Since our local condition 1 involves directly the
Weyl tensor,
it turns out that the Newman-Penrose (NP) formalism
is well-adapted to the proof. However, using such a formalism requires
some care because we are not assuming analyticity of the spacetime and
therefore adapting the null tetrad so that some spin
coefficients vanish requires an existence proof which may be difficult.
So, we will avoid adapting the tetrad except when this existence issue
is easily solved. Second, since the null tetrads in 
the NP formalism are defined only locally (i.e. in a suitable neighbourhood
of each point) we write explicitly the domain of validity of 
the expressions involved (sometimes we indicate this domain in the text
immediately before the formulas).
This makes the exposition more precise but
forces a slightly cumbersome notation.
For the sake of clarity we have collected the various definitions that we
need, even though they may not be used immediately.
These definitions appear at the end of sections 3 and 4 and
they should be looked at for reference.

\vspace{5mm}

\section{Notation, definitions and basic equations.}
\label{sec1}
A $C^n$ spacetime denotes a paracompact, Hausdorff and
connected $C^{n+1}$ four-di\-men\-sio\-nal
manifold endowed with a $C^n$ metric of
signature $(-1,1,1,1)$. Smooth means $C^{\infty}$. In this paper,
all spacetimes are assumed to
be oriented with metric volume form 
$\eta_{\alpha\beta\gamma\delta}$. The conventions and notation for the
Riemann and Ricci tensors follow \cite{HE}.
Throughout this paper  $(\M,g)$ will
denote a spacetime satisfying the hypotheses of Theorem \ref{Main}.
The norm and twist of the Killing vector $\vec{\xi}$
are defined as $\lambda = - \xi^{\alpha} \xi_{\alpha}$ and 
$\omega_{\alpha} = \eta_{\alpha\beta\gamma\delta} \xi^{\beta} \nabla^{\gamma}
\xi^{\delta}$ respectively. As mentioned in the introduction,
it is convenient
to employ self-dual 2-forms, which are complex
2-forms $\cm{B}$ satisfying $\cm{B}^{\star} = -i\cm{B}$,
where $\star$ is the Hodge dual operator. In particular, the 2-form 
$\FF_{\alpha\beta} = \nabla_{\alpha} \xi_{\beta}$ and the so-called 
{\it Killing form} $\F_{\alpha\beta} \equiv
 \FF_{\alpha\beta} + i \FF^{\star}_{\alpha\beta}$ (which is self-dual
by definition) will play a fundamental
role. The Ernst one-form is defined as $\chi_{\mu}
\equiv 2 \xi^{\alpha} \F_{\alpha
\mu} = \nabla_{\mu} \lambda - i \omega_{\mu}$. 

From the Killing equations $\nabla_{\alpha} \xi_{\beta} +
\nabla_{\beta} \xi_{\alpha}=0$ it follows $\nabla_{\mu}
\nabla_{\alpha} \xi_{\beta} = \xi^{\nu} C_{\nu \mu \alpha \beta}$,
where $C_{\alpha\beta\gamma\delta}$ is the Weyl tensor.
Consequently, we have
\begin{eqnarray}
\nabla_{\mu} \F_{\alpha \beta} = \xi^{\nu} \C_{\nu\mu\alpha\beta},
\label{fund}
\end{eqnarray}
where $\C_{\nu\mu\alpha\beta} \equiv C_{\nu\mu\alpha\beta} + \frac{i}{2} 
\eta_{\alpha\beta\rho\sigma} C_{\nu\mu}^{\,\,\,\,\,\,\,\, \rho\sigma}$
is the so-called {\it self-dual
Weyl tensor } from the right. In vacuum, this tensor is a double
symmetric, 2-form, trace-free and satisfies the first and
second Bianchi identities
$\C_{\alpha\left [\beta\gamma \delta \right ]} = 0$,
$\nabla^{\nu} \C_{\nu\mu\alpha\beta}=0$. Using well-known identities
involving self-dual 2-forms (see e.g. \cite{Mars1}
for a brief summary) we easily find
\begin{eqnarray}
 \chi_{\alpha}  \chi^{\alpha} = -\lambda \Fsq,
\hspace{1cm}
\nabla_{\alpha} \chi_{\beta} - \nabla_{\beta} \chi_{\alpha}=0,
\hspace{1cm}
\nabla^{\alpha} \chi_{\alpha} = - \Fsq,
\label{divchi}
\end{eqnarray}
where $\Fsq \equiv \F_{\alpha\beta} \F^{\alpha\beta}$.
The second expression shows that the Ernst
one-form is closed and hence locally exact.
Similar, although longer calculations, allow us to establish the
following two identities, which will be useful in the following
\begin{eqnarray}
\nabla_{\mu} \Fsq = 2 \xi^{\nu} \C_{\nu\mu\alpha\beta}
\F^{\alpha\beta}, \label{idennabF2} \hspace{2cm}\\
\nabla_{\alpha} \nabla^{\alpha} \Fsq=
- \C_{\alpha\beta\lambda\mu} \F^{\alpha\beta} \F^{\lambda\mu}
- \frac{\lambda}{2} \C_{\alpha\beta\lambda\mu}  \C^{\alpha\beta\lambda\mu}. 
\label{idenFC}
\end{eqnarray}
Let us now define the subset
$\hat{\M}_f = \left \{ q \in \M \, ; \Fsq |_q \neq 0
\right \}$. Since by assumption $\vec{\xi}$ tends to a time
translation at infinity, $\vec{\xi}$
can be normalized without loss of generality
so that $\lambda \rightarrow 1$ at the asymptotically flat end $\Minf$.
Then, the asymptotic form of the metric in
$\Minf$ (see Remark 1 after Theorem \ref{Main})
implies $\Fsq = -4 M^2 / r^4 + O \left (r^{-5} \right )$ which
implies that 
$\hat{\M}_f$ is not empty (the Komar mass $M$ is non-zero by
hypothesis). Thus, we can assume (perhaps
after restricting $\Minf$ to a proper subset) that 
$\Minf \subset \hat{\M}_f$ and hence the spacetime
$\left (\Mun, \gun \right )$ can be defined as the connected component of
$\hat{\M}_f$ containing $\Minf$, with the induced metric. 
The scalar product of two
vectors $\vec{u}$ and $\vec{v}$ on $\left (\Mun, \gun \right )$
will be denoted simply by
$(\vec{u},\vec{v} )$.

By definition, the complex function $\Fsq$ does not vanish anywhere on
$\Mun$. Let us consider the eigenvalue problem $\F_{\alpha\beta} V^{\beta}
\propto  V_{\alpha}$ with $V^{\alpha}$ real (so, we are actually 
considering two different eigenvalue problems, namely the real 
and the imaginary parts of the previous one).
At points where $\F_{\alpha\beta}$ is regular (i.e.
$\Fsq \neq 0$) this eigenvalue problem is known to have two simple
eigenvalues $f$ and $-f$ satisfying $f^2 = - \Fsq /4$ (this
being a direct consequence
of $\F_{\mu\beta} \F_{\nu}^{\,\,\,\beta} =
(1/4) g_{\mu\nu} \Fsq$, which is a standard property of a self-dual 2-form).
Moreover, simple roots of a polynomial depend smoothly on the polynomial
coefficients (see e.g. \cite{Hall2}).  
The characteristic polynomial of the eigenvalue problem 
${\bf F}_{\alpha\beta} V^{\beta} \propto  V_{\alpha}$
(resp. ${\bf F^{\star}}_{\alpha\beta}
V^{\beta} \propto V_{\alpha}$ ) is biquadratic and has exactly two
real simple roots (whenever $\Fsq \neq 0$) which correspond to
the real (resp. imaginary) parts of $f$ and $-f$. Consequently $f$ and $-f$
depend smoothly on the
coefficients of $\F_{\alpha\beta}$ and therefore are smooth functions on 
$\M_f$ (this fact is not obvious a priory
because $\M_f$ need not be simply connected and therefore the square root
of $\Fsq$ might not be globally defined in $\M_f$).
Let us also call $\vec{k}$ and $\vec{l}$ the
two corresponding real eigenvectors, i.e. $\F_{\alpha\beta} k^{\beta} =
f k_{\alpha}$ and  $\F_{\alpha\beta} l^{\beta} = -f l_{\alpha}$. These vectors
are null and can be normalized, without loss of generality, so that
$\left ( \vec{l}, \vec{k} \right )= -1$. 
Using the fact that $\F_{\alpha\beta}$ is self-dual we can write
\begin{eqnarray}
\left . \F_{\alpha\beta} \right |_{\Mun} = \left .
f \left ( - k_{\alpha} l_{\beta} +k_{\beta} l_{\alpha}
- i \eta_{\alpha\beta\gamma\delta} k^{\gamma} l^{\delta} \right )
\right |_{\Mun}
\Longrightarrow \\
\left . \chi_{\beta} \right |_{\Mun} = \left .
2 f \left [ - \left (\vec{\xi}, \vec{k} \right) l_{\beta}
+ \left (\vec{\xi}, \vec{l} \right) k_{\beta} 
- i \eta_{\alpha\beta\gamma\delta} \xi^{\alpha} k^{\gamma} l^{\delta} \right ]
\right |_{\Mun}.
\label{chialpha}
\end{eqnarray}
This expression shows, in particular, that $\vec{k}$ and $\vec{l}$ are
smooth vector fields on $\M_f$. A simple consequence of the first
equation, together with the fact that ${\pounds}_{\vec{\xi}\,} \F_{\alpha\beta}
=0$ (where ${\pounds}$ denotes Lie derivative) and $\left ( \vec{k},
\vec{l} \, \right ) = -1$, is
\begin{eqnarray}
\left [ \vec{\xi}, \vec{k} \, \right ]  = \left .
C_1 \vec{k} \, \right |_{\Mun}, 
\hspace{1cm}
 \left [ \vec{\xi}, \vec{l} \, \right ] =  \left .- C_1 \vec{l} 
\, \right |_{\Mun}.
\label{comxik}
\end{eqnarray}
where $C_1$ is a smooth function on $\Mun$.
Let us define the self-dual 2-form 
$\left . \bm{W}_{\alpha\beta} \right |_{\Mun} \equiv
 \left . f^{-1} \F_{\alpha\beta}  
\right |_{\Mun}$ which obviously satisfies $\bm{W}_{\alpha\beta}
\bm{W}^{\alpha\beta} = -4$. The space of self-dual 2-forms at
any point $q \in \Mun$ is a complex three-dimensional vector space. We
can therefore complete $\bm{W} |_q$ with two self-dual 2-forms
$\bm{U} |_q$ and $\bm{V} |_q$ such that $\left . \left (  \bm{U},
\bm{V}, \bm{W} \right) \right |_q$ forms a basis. Furthermore, 
it is well-known  that $\bm{U} |_q$ and $\bm{V} |_q$ can
be chosen so that
$\bm{U}_{\alpha\beta} \bm{V}^{\alpha\beta} |_q = 2$ and all the remaining
contractions vanish. This choice of $\bm{U}$ and $\bm{V}$
can be made
locally smooth (i.e. given any point $q \in \Mun$ there exists an open 
neighbourhood of $q$ where $\bm{U}$ and $\bm{V}$ are smooth)
but this choice may not exist globally on  $\Mun$.
So, all the expressions involving $\bm{U}$ and
$\bm{V}$ should be understood to hold in a neighbourhood of each
point in $\Mun$. Since
$\C_{\alpha\beta\gamma\delta}$ is a double, symmetric, self-dual 2-form, it
can be expanded in this basis. The hypothesis 1 of the
theorem implies the existence of a function $\Psi_2$ on $\M$ defined
by $\C_{\alpha\beta\gamma\delta} \F^{\gamma\delta} = -
8 \Psi_2 \F_{\alpha\beta}$ (i.e. the eigenvalue corresponding to the
eigenvector $\F_{\alpha\beta}$). Since
$8 \Psi_2 \Fsq = -\C_{\alpha\beta\gamma\delta} \F^{\alpha\beta}
\F^{\gamma\delta}$ and $\Fsq$ is non-zero on $\Mun$ it follows that 
$\Psi_2$ is smooth on $\Mun$.
Using the symmetries of 
$\C_{\alpha\beta\gamma\delta}$  we obtain the decomposition
\begin{eqnarray}
\left . \C_{\alpha\beta\gamma\delta} \right |_{\Mun} =
\left . 
 2 \Psi_0 \bm{U}_{\alpha\beta} \bm{U}_{\gamma\delta} 
+ 2 \Psi_4 \bm{V}_{\alpha\beta} \bm{V}_{\gamma\delta}
+ 2 \Psi_2 \left ( \bm{W}_{\alpha\beta} \bm{W}_{\gamma\delta}
+ \bm{U}_{\alpha\beta} \bm{V}_{\gamma\delta} +
\bm{V}_{\alpha\beta} \bm{U}_{\gamma\delta} \right ) \right |_{\Mun},
\label{Decom}
\end{eqnarray}
where  
$\Psi_0$ and $\Psi_4$ are locally smooth complex functions (i.e.
they are smooth on the same neighbourhood where
$\left (\bm{U},\bm{V},\bm{W} \right )$ is a smooth basis). The complex
scalar $\C_{\alpha\beta\gamma\delta}
\C^{\alpha\beta\gamma\delta} |_{\Mun} = 32 
( \Psi_0 \Psi_4 + 3 \Psi_2^2) |_{\Mun}$ is obviously smooth everywhere
and therefore $\Psi_0 \Psi_4$ is also a  global smooth function on $\Mun$.
Let us now write down the identities (\ref{idennabF2}), (\ref{idenFC})
on $\Mun$.  In terms of $f$, (\ref{idennabF2}) becomes simply
\begin{eqnarray}
\left . \nabla_{\alpha} f = \frac{\Psi_2}{f} \chi_{\alpha} \right |_{\Mun}
\label{nabf},
\end{eqnarray}
which implies $\chi_{\left [ \alpha \right .}
\nabla_{\left . \beta \right ]} \Psi_2  |_{\Mun} = 0$ because
$\chi_{\alpha}$ is a closed one-form.
Since $\chi_{\alpha}$ cannot vanish identically
on any non-empty open subset of $\Mun$
(otherwise the third equation in (\ref{divchi})
would imply $\Fsq=0$ which cannot happen on $\Mun$) it follows
$\nabla_{\alpha} \Psi_2 |_{\Mun} =
\left . G_1 \chi_{\alpha} \right |_{\Mun}$ for some smooth complex
 function $G_1$.   Inserting this expression into 
(\ref{idenFC}) a short calculation gives us 
an explicit expression for $G_1$, namely 
$G_1 |_{\Mun} =  (2 f^2)^{-1}  ( \Psi_0 \Psi_4 + 3 \Psi_2^2 )
|_{\Mun}$. It turns out to be  convenient to define
a complex one-form $P_{\alpha} = 
(2f)^{-1} \chi_{\alpha}$ which will be used thoroughly from now on.
Notice that (\ref{nabf}) implies that 
$P_{\alpha}$ is closed and hence locally exact. 
In terms of  $P_{\alpha}$, the expressions discussed above take the form
\begin{eqnarray}
\left .\chi_{\alpha} = 
2 f P_{\alpha} \right |_{\Mun}, \hspace{1cm}
\left .\nabla_{\alpha} f = 2 \Psi_2 P_{\alpha} \right |_{\Mun}, \hspace{1cm}
\left . \nabla_{\alpha} \Psi_2 =  \frac{1}{f} \left ( \Psi_0 \Psi_4
+ 3 \Psi_2^2 \right ) P_{\alpha} \right |_{\Mun} , \nonumber \\
\left . P_{\beta}  = 
 - \left (\vec{\xi}, \vec{k} \right) l_{\beta}
+ \left (\vec{\xi}, \vec{l} \right) k_{\beta} 
- i \eta_{\alpha\beta\gamma\delta} \xi^{\alpha} k^{\gamma} l^{\delta} \,
\right |_{\Mun}. \label{exP}
\hspace{25mm}
\end{eqnarray}
As discussed in the introduction, we will need to work
on different subsets of $\Mun$ to prove the theorem.
The relevant definitions are
\begin{eqnarray}
\M_{1} = \left \{ p \in \Mun ; \, \, \,  \Psi_2 |_p= 0 \right \},
\hspace{2mm}
\M_{2} = \left \{ p \in \Mun ; \, \, \, \Psi_2 |_p\neq 0 \mbox{ and }
\Psi_0 \Psi_4 |_p \neq 0 \right \}, \nonumber \\
\M_{3} = \left \{ p \in \Mun ; \, \, \, 
\Psi_2 |_p\neq 0 \mbox{ and } \Psi_0 \Psi_4 |_p = 0\right \}.
 \hspace{3cm} \label{sets}
\end{eqnarray}
By definition, these three sets are a partition of $\Mun$.
A simple analysis of the Petrov types compatible with the 
Weyl tensor (\ref{Decom}) on $\Mun$ shows that Petrov type III is
impossible. The points where the Petrov type is N are contained
in $\M_{1}$, the points where the Petrov type is I are contained in 
$\M_1 \cup \M_2$ and the Petrov type on $\M_3$ is II (or D wherever
$\Psi_0=\Psi_4=0$).

Given any subset $U$ of $\Mun$ we will denote its interior in
the topology of $\Mun$ by $\stackrel{o}{U}$ (occasionally
we will also use $\mbox{int} \, (U)$). The topological boundary
of $U$ is denoted by $\partial U$.

To conclude this section we introduce a concept that will be
useful later in order to write down several lemmas in a more compact form.
\begin{definition}
An open subset $U \subset \Mun$ will be called {\bf spatially
bounded with respect to $\Minf$}
if there exists an asymptotically flat, open, submanifold
$\haMinf \subset \Minf$ such that $U \cap \haMinf = \emptyset$.
\end{definition}

\section{Excluding Petrov type I and N on $\Mun$.}

Let us start by proving two simple lemmas
\begin{lemma}
Let $(\Mun,\gun)$ be constructed as in section \ref{sec1}.
Then, the set of points $\{ q \in \Mun \, ; \, \left .
\nabla_{\alpha} \lambda \right |_q = 0 \}$
has empty interior.
\label{null}
\end{lemma}
{\it Proof:}  Assume that there exists a non-empty open
set $U \subset \Mun$ where $\nabla_{\alpha} \lambda$ vanishes
identically. From the third identity in  (\ref{divchi}) it follows
that $\Fsq$ is purely imaginary. Then, the first identity
in  (\ref{divchi})
reads $\omega_{\alpha}\omega^{\alpha} |_U = \lambda \Fsq |_U$ which immediately
implies $\left . \lambda  \right |_U =0$ and $\left . \omega_{\alpha}
\omega^{\alpha} \right |_U = 0$. Since $\xi^{\alpha}
\omega_{\alpha} = 0$ everywhere and using the fact that the set of zeros of
a Killing vector has empty interior \cite{KN},
it follows that there exists a
smooth function $S$ on $U$ such that $\omega_{\alpha}
 = S \xi_{\alpha} |_U$. Using $\vec{\xi} \left (S
\right ) = 0$, which  follows from ${\pounds}_{\vec{\xi} \,} \omega_{\alpha}
=0$, the third  identity in (\ref{divchi}) implies
\begin{eqnarray*}
\nabla^{\alpha} \chi_{\alpha} = - \mbox{i} \left (
\xi^{\alpha} \nabla_{\alpha} S + S \nabla_{\alpha} \xi^{\alpha} \right )
= 0 = -\Fsq,
\end{eqnarray*}
which is impossible on $U \subset \Mun. \hfill \Box$

\vspace{5mm}

%
%


\begin{lemma}
Let $(\M,g)$ be a spacetime satisfying the hypotheses of Theorem
\ref{Main} and let $\Mun$ and $\MN$ be defined as
in section \ref{sec1}. Then, the set $\MN$ is spatially bounded with respect
to $\Minf$.
\label{MN}
\end{lemma}
{\it Proof:} $\MN$ is open by definition. If $\MN$ is empty the statement
is trivial. If $\MN$ is non-empty, then equation 
({\ref{nabf}) shows that $f$ and hence $\Fsq$ are constant on each connected
component of $\MN$. As we discussed above,
the asymptotic form of the metric in the asymptotically
flat end $\Minf$ implies  $\Fsq = - 4 M^2/ r^{4} + O (r^{-5})$. Hence,
there exists an asymptotically flat open submanifold $\haMinf$ 
where $\nabla_{\alpha} \Fsq$ is nowhere zero. Thus $\MN \cup  \haMinf =
\emptyset$ and the lemma is proven. $\hfill \Box$.

\vspace{5mm}

Take now an arbitrary point $p \in \Mun$ and choose an open,
connected neighbourhood $U_p$ of $p$ where the basis $(\bm{U},
\bm{V}, \bm{W} )$ can be chosen smoothly. It is well-known
\cite{Israel}
that there exists a smooth complex vector $\vec{m}$ on $U_p$
such that $\left \{ \vec{k}, \vec{l}, \vec{m},
\vec{\ov{m}} \right \}$ (the bar denotes complex conjugate)
is a positively oriented null tetrad 
(i.e. $\eta_{\alpha\beta\gamma\delta} k^{\alpha} l^{\beta} m^{\gamma}
\ov{m}^{\delta} = -i$) and such that
$\bm{U}, \bm{V}$ and $\bm{W}$ take the form
\begin{eqnarray}
\left .\bm{U}_{\alpha\beta} \right |_{U_p}= \left .
- l_{\alpha} \ov{m}_{\beta} +
 l_{\beta} \ov{m}_{\alpha}\right |_{U_p}, \hspace{1cm}
\left .\bm{V}_{\alpha\beta} \right |_{U_p} = \left . k_{\alpha} m_{\beta} -
 k_{\beta} m_{\alpha} \right |_{U_p}, \nonumber \\
\left . \bm{W}_{\alpha\beta} \right |_{U_p}= \left .
- k_{\alpha} l_{\beta} + k_{\beta} l_{\alpha}
+ m_{\alpha} \ov{m}_{\beta} - m_{\beta} \ov{m}_{\alpha} \right |_{U_p}.
\label{UVW}
\end{eqnarray}
We shall use the Newman-Penrose (NP) notation to denote the Ricci rotation
coefficients associated with this null basis. Our conventions will follow
\cite{KSMH}. The names $\Psi_0$, $\Psi_2$ and $\Psi_4$ above 
were chosen precisely so that they correspond to the spin coefficients
of the Weyl tensor. In particular, (\ref{Decom}) shows that
$\Psi_1 = \Psi_3= 0$.
For completeness, let us write down the Newman-Penrose equations. They
are just the standard definition of the curvature in terms of the
connection when everything is expressed
in the null basis $\{ \vec{k}, \vec{l}, \vec{m},
\vec{\ov{m}} \}$ and therefore they only hold on $U_p$
\begin{eqnarray}
D \rho - \ov{\delta} \kappa & = &\rho^2+
\sigma \ov{\sigma}+\rho \left (\epsilon+\ov{\epsilon} \right )
-\ov{\kappa} \tau- \kappa \left (3 \alpha+\ov{\beta}-\pi \right 
) \label{eq728} \\
D \sigma -\delta \kappa & = &\sigma \left (\rho+\ov{\rho} \right 
)+ \left (3 \epsilon-\ov{\epsilon} \right )
\sigma- \left (\tau-\ov{\pi}+\ov{\alpha} +3 \beta \right )
\kappa+\Psi_0 \label{eq729} \\
D \tau -\Delta \kappa & = &
\rho \left (\tau+\ov{\pi} \right )+\sigma \left (\ov{\tau}+\pi
\right )+\tau \left (\epsilon-\ov{\epsilon} \right )-
\kappa \left (3 \gamma+\ov{\gamma} \right )+\Psi_1 \label{eq730} \\
D \alpha -\ov{\delta} \epsilon & = &
\alpha \left (\rho+\ov{\epsilon}-2 \epsilon \right )
+\ov{\sigma} \beta-\ov{\beta} \epsilon-
\kappa \lambda-\ov{\kappa} \gamma+\pi \left (\epsilon+\rho
\right ) \label{eq731} \\
D \beta -\delta \epsilon & = &
\sigma \left (\alpha+\pi \right )+\beta \left (\ov{\rho}-\ov{\epsilon}
\right )
-\kappa \left (\mu+\gamma \right )-\epsilon  \left 
(\ov{\alpha}-\ov{\pi} \right )+\Psi_1 \label{eq732} \\
D \gamma -\Delta \epsilon & = &
\alpha \left (\tau+\ov{\pi} \right )+\beta \left (\ov{\tau}+\pi
\right )-\gamma \left (\epsilon+\ov{\epsilon} \right )
-\epsilon \left (\gamma+\ov{\gamma} \right )+\tau \pi-\nu \kappa+\Psi_2 
\label{eq733} \\
D \lambda-\ov{\delta} \pi& = &
\rho\ \lambda+\ov{\sigma} \mu+\pi^2+\pi \left (\alpha-\ov{\beta} \right )
-\nu \ov{\kappa}-\lambda \left (3 \epsilon-\ov{\epsilon} \right
) \label{eq734} \\
D \mu - \delta \pi & = &
\ov{\rho} \mu+\sigma \lambda+\pi \ov{\pi}-\mu \left (\epsilon+\ov{\epsilon}
\right )
-\pi \left (\ov{\alpha}-\beta \right )-\nu \kappa+\Psi_2 \label{eq735} \\
D \nu -\Delta \pi & = &
\mu \left (\pi+\ov{\tau} \right )+\lambda \left (\ov{\pi}+\tau \right )+
\pi \left (\gamma-\ov{\gamma} \right ) 
-\nu \left (3 \epsilon+\ov{\epsilon} \right )+\Psi_3 \label{eq736} \\
\Delta \lambda - \ov{\delta} \nu & = &-\lambda \left (\mu+\ov{\mu}
\right )-\lambda \left (3
\gamma-\ov{\gamma} \right )+\nu \left (3 \alpha+\ov{\beta} +\pi -\ov{\tau}
\right )-\Psi_4  \label{eq737} \\
\delta \rho - \ov{\delta} \sigma & = &
\rho \left (\ov{\alpha}+\beta \right )-\sigma \left (3 \alpha-
\ov{\beta} \right )
+\tau \left (\rho-\ov{\rho} \right )+\kappa \left (\mu-\ov{\mu}
\right )- \Psi_1 \label{eq738} \\
\delta \alpha - \ov{\delta} \beta & = &
\mu \rho-\lambda \sigma+\alpha \ov{\alpha}+\beta \ov{\beta}-
2 \alpha \beta+\gamma \left (\rho-\ov{\rho} \right )+
\epsilon \left (\mu-\ov{\mu} \right )-\Psi_2 
\label{eq739} \\
\delta \lambda -\ov{\delta} \mu & = &
\nu \left (\rho-\ov{\rho} \right )+\pi \left (\mu-\ov{\mu} \right
)+\mu \left (\alpha+\ov{\beta} \right )
+\lambda \left (\ov{\alpha}-3 \beta \right ) \label{eq740} \\
\delta \nu -\Delta \mu & = &
\mu^2+\lambda \ov{\lambda}+\mu \left (\gamma+\ov{\gamma} \right
)-\ov{\nu} \pi
+\nu \left (\tau-3 \beta-\ov{\alpha} \right ) \label{eq741} \\
\delta \gamma -\Delta \beta & = &
\gamma \left (\tau-\ov{\alpha}-\beta \right )+
\mu \tau-\sigma \nu-\epsilon \ov{\nu}
-\beta \left (\gamma-\ov{\gamma} -\mu \right )+
\alpha \ov{\lambda} \label{eq742} \\
\delta \tau- \Delta \sigma & = &
\mu \sigma+\ov{\lambda} \rho+\tau \left (\tau+\beta-\ov{\alpha}
\right )-
\sigma \left (3 \gamma-\ov{\gamma} \right )-\kappa \ov{\nu}
 \label{eq743} \\
\Delta \rho -\ov{\delta} \tau & = &
-\rho \ov{\mu}-\sigma \lambda+\tau \left (\ov{\beta}-\alpha-\ov{\tau}
\right )+
\rho \left (\gamma+\ov{\gamma} \right )+\nu \kappa-\Psi_2 \label{eq744} \\
\Delta \alpha -\ov{\delta} \gamma & = &
\nu \left (\rho+\epsilon \right )-\lambda \left (\tau+\beta \right )+
\alpha \left (\ov{\gamma}-\ov{\mu} \right )+\gamma \left 
(\ov{\beta}-\ov{\tau} \right )  \label{eq745} 
\end{eqnarray}
Taking into account the third equation in (\ref{exP}), the only non-trivial 
second Bianchi identities for the Weyl tensor can be written on $U_p$ as
\begin{eqnarray}
\ov{\delta} \Psi_0  & = &
\Psi_0 \left (4 \alpha-\pi \right )+
3 \kappa \Psi_2 \label{Bian761}, \\
\Delta \Psi_0  & = &
\Psi_0 \left (4 \gamma-\mu \right ) +3 \sigma \Psi_2 \label{Bian762}, \\
 - D \Psi_4 & = &
\Psi_4 \left (4 \epsilon-\rho \right )
+3 \lambda \Psi_2 \label{Bian763} ,\\
 - \delta \Psi_4 & = &
\Psi_4 \left (4 \beta-\tau \right )+3 \nu \Psi_2 \label{Bian764}, 
\end{eqnarray}
while the commutators on $U_p$ between the vectors $\left \{ \vec{k},
\vec{l}, \vec{m}, \vec{\ov{m}} \right \}$ read
\begin{eqnarray}
\Delta D - D \Delta  & = & \left (\gamma+\ov{\gamma}
\right ) D + \left (\epsilon+\ov{\epsilon} \right ) \Delta 
- \left (\tau+\ov{\pi} \right ) \ov{\delta} -
\left (\ov{\tau}+\pi \right ) \delta \label{commDelD} ,\\
\delta D  - D \delta  & = & \left (\ov{\alpha}+\beta-\ov{\pi} \right ) D 
+\kappa \Delta -\sigma \ov{\delta} - \left 
(\ov{\rho}+\epsilon-\ov{\epsilon} \right ) \delta \label{commdelD}, \\
\delta \Delta  - \Delta \delta & = &  -\ov{\nu} D +\left 
(\tau-\ov{\alpha}-\beta \right ) \Delta 
+\ov{\lambda} \ov{\delta} + \left (\mu-\gamma+\ov{\gamma} \right
) \delta \label{commdelDel}, \\
\ov{\delta} \delta - \delta \ov{\delta} & = &   \left 
(\ov{\mu}-\mu \right ) D + \left (\ov{\rho}-\rho \right ) \Delta 
- \left (\ov{\alpha}-\beta \right ) \ov{\delta} - \left (\ov{\beta}-\alpha
\right ) \delta \label{commdeldel}. 
\end{eqnarray}
Let us now exploit the identity (\ref{fund}) in order to obtain
information on the NP spin coefficients. In order to do that, we need
to know the covariant derivative of the self-dual
2-form $\bm{W}_{\alpha\beta}$.
The following identity follows directly from the definition of the
NP spin coefficients in any spacetime
\begin{eqnarray*}
\nabla_{\mu} \bm{W}_{\alpha\beta} \equiv 
\left . 2 \left ( \nu k_{\mu} + \pi l_{\mu} - \mu \ov{m}_{\mu}
- \lambda m_{\mu} \right ) \bm{V}_{\alpha\beta} + 2 \left (
- \kappa l_{\mu} - \tau k_{\mu} + \rho m_{\mu} + \sigma \ov{m}_{\mu}
\right ) \bm{U}_{\alpha\beta}  \right |_{U_p}.
\end{eqnarray*}
Inserting this identity into equation (\ref{fund}), 
it is straightforward to find the following
expressions (which hold on $U_p$)
\begin{eqnarray}
\nu =  \frac{ - \left ( \vec{\xi}, \vec{m} \right ) \Psi_4}{f}, \hspace{5mm}
\pi =  \frac{ \left ( \vec{\xi}, \vec{\ov{m}} \right ) \Psi_2}{f}, \hspace{5mm}
\mu =  \frac{ \left ( \vec{\xi}, \vec{l} \, \right ) \Psi_2}{f}, \hspace{5mm} 
\lambda =  \frac{ - \left ( \vec{\xi}, \vec{k} \, \right ) \Psi_4}{f}, 
\nonumber \\
\kappa =  \frac{ - \left ( \vec{\xi}, \vec{\ov{m}} \right ) \Psi_0}{f},
\hspace{5mm}
\tau =  \frac{ \left ( \vec{\xi}, \vec{m} \right ) \Psi_2}{f},
\hspace{5mm} 
\rho =  \frac{ \left ( \vec{\xi}, \vec{k} \, \right ) \Psi_2}{f},
\hspace{5mm}
\sigma =  \frac{ - \left ( \vec{\xi}, \vec{l} \, \right ) \Psi_0}{f}. 
\label{NPcoeff}
\end{eqnarray}
These equations simplify the set of NP equation because the number
of unknowns is reduced considerably. Furthermore
the condition $\left ( \vec{\xi}, \vec{\xi} \, \right ) = -\lambda$
can be rewritten in the null basis as
\begin{eqnarray}
\left .\left ( \vec{\xi}, \vec{k} \,\right ) 
\left ( \vec{\xi}, \vec{l} \, \right )  -
\left ( \vec{\xi}, \vec{m} \right ) 
\left ( \vec{\xi}, \vec{\ov{m}} \right ) = \frac{\lambda}{2} \right |_{U_p} \label{rel}.
\end{eqnarray}
In Lemma \ref{MN} we showed that the interior of the set $\M_1$ cannot
extend to infinity. Our next aim is to show that the same happens
for $\M_{2}$. 
\begin{lemma}
Let $(\M,g)$ be a spacetime satisfying the hypotheses of Theorem
\ref{Main} and let $\Mun$ and $\M_2$ be defined as
in section \ref{sec1}. Then, $\M_2$ is spatially bounded with respect
to $\Minf$.
\label{M2}
\end{lemma}
{\it Proof:} $\M_2$ is open because $\Psi_2$ 
and $\Psi_0 \Psi_4$ are global smooth functions
on $\Mun$. If $\M_2$ is empty the
proposition is trivial, so let us assume that $\M_2$ is non-empty 
and that intersects every asymptotically flat open submanifold $\haMinf$.
An immediate consequence of (\ref{exP}) is the existence of
a complex function $G_2$ on $\Mun$ such that $\nabla_{\alpha}
\left (\Psi_0 \Psi_4 \right ) =
G_2 P_{\alpha}$. Using this expression together
with the  Bianchi identities (\ref{Bian763}) and (\ref{Bian764})
and equation (\ref{rel}), the combination
$\Psi_0 \times (\ref{eq735}) + \Psi_4 \times (\ref{eq729})$ 
takes the simple form
\begin{eqnarray}
\left . \left ( \frac{G_2}{4 f} - \frac{2 \Psi_0 \Psi_4 \Psi_2}{f^2} \right )
\lambda  \right |_{\MI} = \left . \frac{}{} - \Psi_0 \Psi_4  \right |_{\MI}.
\label{chichibar}
\end{eqnarray}
This equation implies, first of all, that neither
$ G_2 f  - 8  \Psi_0 \Psi_4 \Psi_2 $ nor $\lambda$ 
can vanish anywhere on $\MI$. Then, taking the gradient of
(\ref{chichibar}) and using $\nabla_{\alpha} \lambda = 1/2
(\chi_{\alpha} + \ov{\chi}_{\alpha})$ (which follows from the definition of
the Ernst one-form) and the fact that $\nabla_{\alpha} G_2 \propto
P_{\alpha}$ (which follows from the definition of $G_2$) we obtain
\begin{eqnarray*}
\left . \left ( G_2 f  - 8 \Psi_0 \Psi_4 \Psi_2 \right )
\chi_{\left [ \alpha \right .}  \ov{\chi}_{\left . \beta \right ]}
\right |_{\MI} = 0 \hspace{1cm}
\Longrightarrow 
\hspace{1cm}
\left .
\omega_{\left [ \beta \right .} \nabla_{\left . \alpha \right ]} \lambda 
\right |_{\MI} =0.
\end{eqnarray*}
From Lemma {\ref{null} we have that $\nabla_{\alpha} \lambda$ cannot 
vanish on any open subset of $\MI$ and hence there exists a real
function $L$ such that $\omega_{\alpha} = L \nabla_{\alpha} \lambda$ on $\MI$.
The identity $\chi_{\alpha} \chi^{\alpha} = - \lambda \Fsq$ becomes
\begin{eqnarray}
\left . 4 f^2 \lambda  ( 1 - \mbox{i} L )^{-2} \right |_{\MI}=
\left . \nabla_{\alpha} \lambda \nabla^{\alpha} \lambda \right |_{\MI}.
\label{r2}
\end{eqnarray}
Thus, $4 f^2 \left ( 1 - \mbox{i} L \right )^{-2}$
is real and, since it cannot vanish anywhere (because
 $f \neq 0$ on $\Mun$),
its sign must remain constant on any connected component of $\MI$.
Those connected components where
$4 f^2 \left ( 1 - \mbox{i} L \right )^{-2} < 0$ are obviously spatially
bounded with respect to $\Minf$
because $\lambda$ is necessarily negative there (if
$\lambda > 0$ then $\nabla_{\alpha} \lambda$ is spacelike -- from
$\xi^{\alpha} \nabla_{\alpha}
\lambda = 0$ -- and (\ref{r2}) implies
$4 f^2 \left ( 1 - \mbox{i} L \right )^{-2}
>0$). Thus we only need to consider those connected components 
$\MI^{\{\alpha \}}$ where  $4 f^2 \left ( 1 - \mbox{i} L \right )^{-2} > 0$
(i.e. where  $2 f \left (  1 - \mbox{i } L \right )^{-1}$ is real). 
Equation (\ref{chialpha}) reads, on $\MI$
\begin{eqnarray*}
\left . \nabla_{\beta} \lambda \right |_{\MI} = \left .
\frac{2 f}{\left ( 1 - \mbox{i} L \right )} 
 \left [ - \left (\vec{\xi}, \vec{k} \right) l_{\beta}
+ \left (\vec{\xi}, \vec{l} \right) k_{\beta} 
- i \eta_{\alpha\beta\gamma\delta} \xi^{\alpha} k^{\gamma} l^{\delta} \right ]
\right |_{\MI}.
\end{eqnarray*}
The imaginary part of this equation tells us that,  at every point
on $\MI^{\{\alpha \}}$, $\vec{\xi}$ lies
on the two-plane generated by $\vec{k}$ and $\vec{l}$ and therefore
(\ref{NPcoeff}) $\nu = \pi = \kappa = \tau = 0$.
On $\MI^{\{\alpha \}}$, the one-form $P_{\alpha}$
defined in (\ref{exP}) takes the
form $P_{\alpha} = \left ( 1 - \mbox{i} L \right ) \left (2 f \right )^{-1}
\nabla_{\alpha} \lambda$ and therefore is real. Consider now
an arbitrary point $p \in \MI^{\{\alpha \}}$ and consider an open, simply
connected
neighbourhood of $U_p$ of $p$ in $\MI^{\{\alpha \}}$ such that the basis
$\left \{ \bm{U}, \bm{V}, \bm{W} \right \}$ is well-defined.
We have the freedom to rescale $\bm{U}$ and
$\bm{V}$ according to $\bm{U'} = Q \bm{U}$
$\bm{V'} = Q^{-1} \bm{V}$ where $Q$ is a non-zero, complex, smooth
function defined on $U_p$. This transformation does not change
any of the properties we have used of the 
basis $\left \{ \bm{U}, \bm{V}, \bm{W} \right \}$ and therefore all the
equations above still hold for $\left \{ \bm{U'}, \bm{V'}, \bm{W} \right \}$
if all quantities are substituted by primes (those which are
invariant under this transformation will be written without primes).
Using (\ref{UVW}) we find that
the corresponding vectors $\vec{k}'$, $\vec{l}'$ and $\vec{m}'$ are
related to $\vec{k}$, $\vec{l}$ and $\vec{m}$ by
\begin{eqnarray*}
\vec{k}' = \frac{1}{|Q|} \vec{k}, \hspace{5mm}
\vec{l}' = |Q| \vec{l}, \hspace{5mm} \vec{m}' = e^{- i \varphi} \vec{m},
\hspace{1cm} \mbox{where} \hspace{5mm} Q = |Q| e^{i \varphi}.
\end{eqnarray*} 
Under such a
transformation,  $\Psi_0$ and 
$\Psi_4$ become $\Psi_4'= Q^{2} \Psi_4$
and $\Psi_0' = Q^{-2} \Psi_0$. Since  $\Psi_4$ and $\Psi_0$ are nowhere
vanishing on $U_p$, there exists a smooth
choice of $Q$ such that $\Psi_0'
= \Psi_4'  |_{U_p}$. This choice will be assumed from now on. 
Now, equation (\ref{chichibar}) allows us to obtain $G_2$ explicitly
\begin{eqnarray*}
\left . G_2 \frac{}{} \right |_{U_p} = \left .
\frac{4 \Psi_4'^2 \left (2 \Psi_2 \lambda - f^2 \right )}{f \lambda}
\right |_{U_p}
\hspace{5mm} \Longrightarrow \hspace{5mm} 
\left . \nabla_{\alpha} \Psi_4' \frac{}{} \right |_{U_p} =
\left .  \frac{ 2 \Psi_4' \left (2 \Psi_2 \lambda -  f^2
\right )}{f \lambda} P_{\alpha}
\right  |_{U_p}.
\end{eqnarray*}
The Bianchi equations (\ref{Bian761})-(\ref{Bian764}) immediately imply 
(recall that, on $\MI^{\{
\alpha \} }$,  $\vec{\xi}$ is a non-lightlike linear combination
of $\vec{k}'$ and $\vec{l}'$ from which 
$( \vec{\xi}, \vec{k}' \, ) \neq 0$ and $( \vec{\xi},
\vec{l}' \, )\neq 0$)
\begin{eqnarray*}
\left. \alpha' \right |_{U_p} = \left . \beta' 
\right |_{U_p} = 0, \hspace{1cm}
\left .\gamma' \right |_{U_p} = \left .
 \frac{f}{4 \left (\vec{\xi}, \vec{k}' \, \right )}
\right |_{U_p},
\hspace{1cm}
\left .\epsilon'  \right |_{U_p} = \left .
\frac{f}{4 \left (\vec{\xi}, \vec{l}' \, \right )} \right |_{U_p}.
\end{eqnarray*}
Combining  equations (\ref{eq728}), (\ref{eq733})
and (\ref{eq741})  in order to eliminate $D \left (
\vec{\xi} , \vec{k}' \, \right )$ and $\Delta \left (
\vec{\xi} , \vec{l}' \, \right )$ we obtain,
after dropping some non-zero factors,
\begin{eqnarray*}
 \left . \left (  \ov{\Psi}_4' f \left [
\left (\vec{\xi} , \vec{k}' \, \right )^4
+\left (\vec{\xi} , \vec{l}' \, \right )^4 \right ]
- 2 \ov{f} \Psi_4' \left (\vec{\xi} , \vec{k}' \, \right )^2
\left (\vec{\xi} , \vec{l}' \, \right )^2 \right ) \right |_{U_p}=0,
\end{eqnarray*}
which is equivalent to 
\begin{eqnarray}
\left .\left (\vec{\xi} , \vec{k}' \, \right )^2 \right |_{U_p}= 
\left .\left (\vec{\xi} , \vec{l}' \, \right )^2 \right |_{U_p} , \hspace{1cm}
\left . f \ov{\Psi}_4' |_{U_p} = \ov{f} \Psi_4' \right |_{U_p}
\label{psi4f}
\end{eqnarray}
and therefore  $\left . \left (\vec{\xi} ,\vec{k}' \, \right ) \right
|_{U_p} =  \left . \pm |\lambda/2 | ^{1/2} \right |_{U_p}$ and
$\left . \left (\vec{\xi} , \vec{l}' \, \right ) \right |_{U_p} =
\left .\epsilon_1 
\left (\vec{\xi} , \vec{k}' \, \right ) \right |_{U_p}$, where
$\epsilon_1 = \mbox{sign} (\lambda)$. 
Let us now define four functions $A_i$ ($i = 1,2,3,4$) 
on $U_p$ by the following expressions
\begin{eqnarray}
f \equiv  \frac{\lambda}{2} \left (A_1 + i A_2 \right ), \hspace{1cm}
\Psi_0' = \Psi_4' \equiv  f A_3, \hspace{1cm} \Psi_2 \equiv  f \left (A_4
+ i \frac{A_2}{2} \right ).
\label{defsA}
\end{eqnarray}
$A_1$ and $A_2$ are real by definition, $A_3$ is also real from
(\ref{psi4f}). Furthermore, the NP equation (\ref{eq744}) becomes simply 
$A_4 - \ov{A_4} = 0$ and therefore all $A_i$ are real.
The fact that $2f (1- i L )^{-1}$ is also
real implies that $A_1 |_{U_p}$ is nowhere
zero and that $L  = - A_2 A_1^{-1} |_{U_p}$. It is now a matter of simple,
if somewhat long, calculation to check that the full system of Newman-Penrose
equations (\ref{eq728})-(\ref{eq745}) are satisfied on $U_p$
if and only if
\begin{eqnarray}
\nabla_{\alpha} A_2 = 2 A_2 A_4  P_{\alpha}, \hspace{1cm}
\nabla_{\alpha} A_1 = \left ( 2 A_1 A_4 - A_1^2 - A_2^2 \right ) 
 P_{\alpha}, 
\label{eqsA1A2} \\
\nabla_{\alpha} A_3  = A_3 \left ( 2 A_4 - A_1 \right ) P_{\alpha}, 
\hspace{1cm} 
\nabla_{\alpha} A_4  = \left ( A_3^2 + A_4^2 - \frac{1}{4} A_2^2 \right )
P_{\alpha}, \label{eqsA3A4} \\
\omega_{\alpha} = - A_2 \lambda  P_{\alpha}, \hspace{1cm}
\nabla_{\alpha} \lambda = \lambda A_1 P_{\alpha},  \hspace{2cm}
\label{nalam} \\
4 \left ( A_4^2 - A_3^2 \right ) - 4 A_4 A_1 - A_2^2 = 0. \label{constraint}
\hspace{2cm} 
\end{eqnarray}
Let us now assume that $\M_2$ is not spatially bounded with respect
to $\Minf$. Then it must intersect any asymptotically flat end
$\haMinf$ and we can choose $U_p$ so that $U_p \subset \haMinf$. 
Thus, we can coordinate $U_p$ with the asymptotically Minkowskian
coordinates described in Remark 1 after Theorem {\ref{Main}. We know
from asymptotic flatness that $\Fsq = -4 M^2 / r^4 + O(r^{-5})$ and
therefore $f = \tilde{\epsilon} M /r^2 + O(r^{-3})$
where $\tilde{\epsilon} = \pm 1$. Using the
fact that $\lambda = 1 - 2M/r + O(r^{-2})$ we deduce from the definition
of $A_1$ and $A_2$ in (\ref{defsA}) that
\begin{eqnarray*}
A_1 = \frac{2M\tilde{\epsilon}}{r^2} + O(r^{-3}),
\hspace{2cm}
A_2 = O(r^{-3})
\end{eqnarray*}
Regarding $A_4$, we can obtain its asymptotic behavior directly 
from the second equation in (\ref{eqsA1A2}) after using the second equation in
(\ref{nalam})
and the fact that $A_1$ is non-zero on $U_p$.
We find $A_4 = -\tilde{\epsilon}/r + O(r^{-2})$. Finally,
the constraint (\ref{constraint}) fixes the asymptotic behaviour
of $A_3$ which is $A_3 = \hat{\epsilon}/r 
+ O (r^{-2}) $ where $\hat{\epsilon} = \pm 1$. A direct consequence
of the first equation in (\ref{eqsA3A4}) and the second one in
(\ref{nalam}) is
\begin{eqnarray*}
\lambda A_1 \nabla_{\alpha} A_3 - A_3 \left (2 A_4 - A_1 \right )
\nabla_{\alpha} \lambda = 0.
\end{eqnarray*}
Inserting the asymptotic expansions  into this equation,
we obtain $2 M \hat{\epsilon} \tilde{\epsilon}/r^{4} + O(r^{-5}) = 0$.
Choosing $U_p$ so that $r$ takes values which are large enough this
identity becomes impossible.  Thus $\M_2$ must be spatially bounded
with respect to $\Minf$. $\hfill \Box$

\vspace{5mm}

The main result in this section is 
Proposition \ref{Mun=M3} below. Before proving it,
we must discuss some of the basic
properties of the set $\M_3$ (where the Petrov type is II
or D ).
In the following section we will 
study $\M_3$ in detail, but, for now, 
we only need to notice that inserting $\Psi_0 \Psi_4 = 0$ and
$\Psi_2 \neq 0$ (which define $\M_3$) into
(\ref{exP}) we get  $\nabla_{\beta} \Psi_2 |_{\M_3} = (3/2) \Psi_2 f^{-1}
\nabla_{\beta} f |_{\M_3}$. This implies immediately that
$\Psi_2^2 / f^3$ is constant on any connected component of $\M_3$. 
Since $\Psi_2$ is a smooth, globally defined complex function on $\M_3$,
it follows easily that $f^{1/2}$ defines a smooth, global function on
$\M_3$. Thus we can write
\begin{eqnarray}
\left . \Psi_2 \right |_{\M_3} = - \left . \frac{\sqrt{2}}{b}
\left ( f^{1/2} \right )^3 \, \,\right |_{\M_3},
\label{psi2}
\end{eqnarray}
where $b$ is a nowhere zero, complex function on $\M_3$ which is constant
on each connected component of $\M_3$.

\begin{proposition}
Let $(\M,g)$ be a spacetime satisfying the hypotheses of Theorem
\ref{Main} and let $\Mun$, $\M_1$, $\M_2$ and $\M_3$ be defined as
in section \ref{sec1}.
Then $\M_f = \M_3$.
\label{Mun=M3}
\end{proposition}
{\it Proof:} Trivial topological properties of 
$\M_1$, $\M_2$ and $\M_3$ together with the fact that they constitute
a  partition of $\Mun$ show that $\partial ( \M_1 \cup \M_2 ) =
\partial \M_3$. Moreover, it is clear that $\partial (\M_1 \cup \M_2 )
\subset \partial \M_1 \cup \partial \M_2$ and therefore we have
\begin{eqnarray*}
\partial \M_3 = \partial \M_3 \cap \partial \left (  \M_1 \cup \M_2 
\right )
\subset \partial \M_3 \cap \left ( \partial \M_1 \cup \partial \M_2
\right ) = \hspace{2cm} \\
\hspace{4cm} \left ( \partial \M_3 \cap \partial \M_1 \right ) \cup
\left ( \partial \M_3 \cap \partial \M_2 \right ) .
\end{eqnarray*}
So, if we prove $\left ( \partial \M_3 \cap \partial \M_1 \right ) = 
\left ( \partial \M_3 \cap \partial \M_2 \right ) = \emptyset$ then
we would have
$\partial \M_3 = \emptyset$ which implies easily $\M_3 = \Mun$ (after
using Lemmas \ref{MN} and \ref{M2} above).
So, our aim is to show that both 
$\partial \M_3 \cap \partial \M_1$ and 
$\partial \M_3 \cap \partial \M_2$ are empty. 
Showing  $\partial \M_3 \cap \partial \M_1 = \emptyset$ is trivial because
at any point $q \in \partial \M_3 \cap \partial \M_1$ we would have
$\Psi_2 |_q =0$ from the continuity of $\Psi_2$ and the definition of $\M_2$
and this is incompatible with the limit of $\Psi_2$ on $q$ coming
from $\M_3$ which is nonzero due to expression (\ref{psi2}) after using the
constancy of $b$ on any connected component of $\M_3$ and the
fact that $f$ cannot become zero anywhere on $\Mun$. So, we only need
to show that $\partial \M_3 \cap \partial \M_2 = \emptyset$. 
Assume, on the contrary, that this set is non-empty
and take an arbitrary  point $q \in \partial \M_3 \cap \partial \M_2$ and
a sufficiently small open, connected neighbourhood $U_q$ of $q$ such
that the null tetrad $\{ \vec{k}, \vec{l}, \vec{m}, \vec{\ov{m}} \}$
is well-defined  on $U_q$. In particular, $\Psi_0$ and $\Psi_4$ are smooth
functions on $U_q$. Let us now find an equation
for the function $\Psi_0\Psi_4 \lambda^2 f^{-2}$. On
$U_q \cap \M_2$ we know that the tetrad $\{\vec{k}',\vec{l}',\vec{m}',
\vec{\ov{m}}' \}$ introduced in the proof of the previous proposition is
well-defined. From (\ref{defsA}) and (\ref{nalam}) and the fact
that $\Psi_0 \Psi_4 = \Psi_0' \Psi_4'$, we  obtain
\begin{eqnarray}
\left .
\nabla_{\alpha} \left ( \frac{\Psi_0 \Psi_4 \lambda^2}{f^2} \right )
= \nabla_{\alpha} \left ( A_3^2 \lambda^2 \right ) =
4 A_3^2 \lambda^2 A_4 P_{\alpha} = 4 \left ( \frac{\Psi_0 \Psi_4 
\lambda^2}{f^2} \right ) 
\mbox{Re} \left ( \frac{\Psi_2}{f} \right ) P_{\alpha}
\right |_{U_q \cap \M_2} ,
\label{eqp0p4}
\end{eqnarray}
where $\mbox{Re}$ denotes real part. 
Take now a smooth curve $\gamma$ entirely contained in $U_q \cap \M_2$
and  with endpoint on $q$ and such that $\lambda$ is non-zero at least
on one point $r$ of $\gamma$. Since $\lambda$ cannot vanish on any open set
(from Lemma \ref{null})  the existence of such a curve is trivial to establish.
Let us parametrize the curve $\gamma$ such that $\gamma(0)=r$ and
$\gamma(1) = q$. Integrating equation (\ref{eqp0p4})
along $\gamma$ we obtain
\begin{eqnarray*}
\frac{\Psi_0 \Psi_4 \lambda}{f^2} (s') = 
\frac{\Psi_0 \Psi_4 \lambda}{f^2} (0) 
\exp \left (\int_{0}^{s'} \left .\left ( 4 \mbox{Re}
\left ( \frac{\Psi_2}{f} \right )
P_{\alpha} \dot{\gamma}^{\alpha} \right  ) \right |_{s} ds \right ), 
\end{eqnarray*}
where $\vec{\dot{\gamma}}$ is the tangent vector of $\gamma$.
Taking into account that 
$ \mbox{Re} \left ( \Psi_2 f^{-1} \right )$ remains bounded along $\gamma$
we conclude that $\Psi_0 \Psi_4 \lambda^2 f^{-2}$ cannot vanish on $q$
which is incompatible with the fact that $\Psi_0 \Psi_4$ vanishes on $q$.
Thus, we have proven that $\Omega = \emptyset$ and the proof is completed.
$\hfill \Box$

\section{Analysis of the Petrov types II or D and proof of the theorem.}

In this section we will concentrate on the subset $\M_3$. From the
previous section we know that $\M_3 = \Mun$ and therefore $\M_3$
is open and
connected. Thus the function $b$ appearing in (\ref{psi2}) 
is constant throughout $\M_3$. The decomposition of
$b$ into real and imaginary parts will be written as $b = b_1 + i b_2$.
Equation (\ref{psi2}) gives the form of $\Psi_2$ on $\M_3$. 
Inserting it into the second equation in
(\ref{exP}) we obtain $\nabla_{\alpha} f = - 2 \sqrt{2} b^{-1}
(f^{1/2})^3 P_{\alpha}$. This equation implies first of all 
that $P_{\alpha}$ is exact on $\M_3$ and that $P_{\alpha} = \nabla_{\alpha} P$
with $P$ given by
\begin{eqnarray}
\left . P = \frac{b}{\sqrt{2}  f^{1/2}} 
\right |_{\M_3} \hspace{1cm} \Longrightarrow
\hspace{1cm} \left . \Psi_2 = - \frac{b^2}{2 P^3} \right |_{\M_3}. 
\label{P}
\end{eqnarray}
Similarly,  the first equation in (\ref{exP}) shows that the Ernst one-form 
$\chi_{\alpha}$ is also exact on $\M_3$ and that $\chi_{\alpha} =
\nabla_{\alpha} \chi$ with 
\begin{eqnarray}
\chi = \left . - \frac{b^2}{P} + c \,\,\right |_{\M_3},
\label{chi}
\end{eqnarray}
where $c$ is a complex constant. Since the imaginary part of the Ernst
potential (i.e. the twist potential) is  defined up to an arbitrary
additive constant, we can assume without loss of generality that $c$ is real.
Let us also define two smooth real functions $y$ and $Z$ on $\M_3$ by 
\begin{eqnarray}
P = y + i Z, \label{defyZ}
\end{eqnarray}
so that (\ref{exP}) becomes
\begin{eqnarray}
\left .\nabla_{\beta} y \right |_{\M_3}  = \left .
 - \left (\vec{\xi}, \vec{k} \right) l_{\beta}
+ \left (\vec{\xi}, \vec{l} \right) k_{\beta} \right |_{\M_3}, \hspace{1cm}
\left .\nabla_{\beta} Z \right |_{\M_3}  = \left .
- \eta_{\alpha\beta\gamma\delta} \xi^{\alpha} k^{\gamma} l^{\delta} \,
\right |_{\M_3}. 
\label{exyZ}
\end{eqnarray}

In order to proceed, we still have to define the following 
subsets of $\M_3$. 
\begin{eqnarray}
\Mk  &= & \mbox{int} 
\left \{ p \in \M_3 ; \, \, \left ( \vec{\xi}, \vec{k} \, \right
 ) |_p \neq 0 \mbox{ and } \Psi_0 |_p= 0 \, \, \right \}, \nonumber \\
\Mpi &=& \left \{ p \in \M_3 ; \, \, \eta_{\alpha\beta\lambda\mu}
\xi^{\beta} k^{\lambda} l^{\mu}|_p \neq 0 \right \}, \nonumber \\
\Ml &= & \mbox{int} 
\left \{ p \in \M_3 ; \, \, \left ( \vec{\xi}, \vec{l} \, \right
 ) |_p \neq 0 \mbox{ and } \Psi_4 |_p = 0 \, \, \right \}, \nonumber \\
\MD  & = &\mbox{int} \,
\left \{ p \in \M_3 ; \, \, \Psi_0 |_p = \Psi_4 |_p = 0 \right \},
\label{defset}  \\
\MPq & = & \left \{ p \in \M_3 ; \, \, \Psi_4 |_p \neq  0 \right \}, 
\hspace{1cm}
\MPz = \left \{ p \in \M_3 ; \, \, \Psi_0 |_p \neq  0 \right \}.
\nonumber
\end{eqnarray}
For later convenience,
let us now write down two equations that hold locally on $\Mk$
(i.e. on a sufficiently small neighbourhood of any
point of $\Mk$). The first one is obtained directly from (\ref{eq734}) after
using $\Psi_0= 0$ and (\ref{eq728}), and the second one follows directly
from (\ref{eq730})
\begin{eqnarray}
\left .
\Psi_4 \left ( \vec{\xi}, \vec{k} \right )^2 \right |_{\Mk}
 = \left .\frac{b^2}{2 P^2}
\left [ \frac{}{}
\left ( \ov{\beta} - \alpha \right ) \left ( \vec{\xi}, \vec{\ov{m}} \right )
- \ov{\delta}  \left ( \vec{\xi}, \vec{\ov{m}} \right ) \right ]
\right |_{\Mk}, 
\label{expsi4}\\
\left .
D \left [ (\vec{\xi}, \vec{m}) \ov{P} \right ] \right |_{\Mk}
 = \left . (\vec{\xi}, \vec{m}) \ov{P}
\left (\epsilon - \ov{\epsilon} \right ) \right |_{\Mk}.
\label{Dxim}
\end{eqnarray}
The following proposition gives the local form of the metric in a certain
subset of $\M_3$. 
As mentioned in the introduction,
asymptotic flatness is not used anywhere in the proof of this proposition.

\begin{proposition}
Let $(S,\gamma)$ be a 
two-dimensional Riemannian space of constant
curvature equal to $2 c$ and
denote by $<\, , \, >$ the scalar product with respect to $\gamma$.
Denote also by $\Delta_{\gamma}$ the Laplace-Beltrami operator on $(S,\gamma)$
and by $\star_{\gamma}$ the Hodge dual of $(S,\gamma)$. Finally, let
$V = \Bbb{R} \times \Bbb{R} \times S$ and $\{u,y\}$
Cartesian coordinates of  $\Bbb{R} \times \Bbb{R}$.

Assume that $\Mpi \cap (\Mk \cup \Ml) $ defined above is non-empty and let
$X$ be a connected component of this set. Then,
there exist three smooth real functions $Z$, $A$ and $B$
defined on $S$ satisfying the linear partial
differential equations 
\begin{eqnarray}
\Delta_{\gamma} Z = -2 c Z + 2 b_1 b_2 , \hspace{2cm}
\label{delZ} \\
\left < dZ, dA \right > - \left <  \star_{\gamma} dZ , d B \right > + 2 Z
- A \left ( 2 c Z - 2 b_1 b_2 \right ) =0,
\label{equaAB}
\end{eqnarray}
($b_1$ and $b_2$ are the two constants defined above)
such that
$(X, g|_X)$ is locally isometric to the spacetime
$(V, h )$ where
\begin{eqnarray}
h  = - \left (
c - \frac{ \left ( b_1^2 - b_2^2 \right ) y + 2 b_1 b_2 Z }{y^2 + Z^2}
 \right ) \left [ \frac{}{}  du - B dZ - A \star_\gamma dZ 
\right ] ^2 + \nonumber \\
+  2 \left ( dy -  \star_\gamma dZ \right )
 \left ( du - B dZ - A \star_\gamma dZ \right )
+ \left (y^2 + Z^2 \right ) \gamma.
\label{metric}
\end{eqnarray}
\label{funda}
\end{proposition}
{\bf Remark.} In the line-element (\ref{metric}) the objects $Z$,
$dZ$, $\star_\gamma dZ$ and $\gamma$ denote the pull-back with
respect to the canonical projection $\hat{\pi} : \Bbb{R} \times \Bbb{R}
\times S \rightarrow S$ of the same objects on $S$. A precise notation should
require giving different names to those pull-backs, but for the sake of
notational simplicity we have preferred not to do so.

\vspace{5mm}

{\it Proof:} Let $p \in X$ be an arbitrary point and consider, as usual,
an open, connected
neighbourhood $U_p \subset X $ of $p$ such that  
null tetrad $\left \{ \vec{k}, \vec{l}, \vec{m}, \vec{\ov{m}} \right \}$
is well-defined on $U_p$. Define $\pi_0$ on $U_p$ as 
$\pi_0 \equiv \left (\vec{\xi},\vec{m} \right ) \ov{P}$. From the definition
of $\Mpi$ we have that $\pi_0$ is nowhere vanishing on $U_p$.
In addition, from $X \subset \Mk \cup \Ml$ it follows that,
by choosing $U_p$ small enough, we can assume that
either $\left (\vec{\xi},\vec{k}\right) \neq 0$ and $\Psi_0=0$ everywhere on
$U_p$ or $\left (\vec{\xi},\vec{l}\right) \neq 0$ and $\Psi_4=0$ everywhere
on $U_p$. We will make the proof explicitly only 
for the case $\left (\vec{\xi},\vec{k}\right) \neq 0$ and $\Psi_0=0$ on $U_p$.
The proof in the other case is identical step by step 
after interchanging the roles of $\vec{k}
\leftrightarrow \vec{l}$, $\vec{m} \leftrightarrow
\vec{\ov{m}}$ and $\Psi_0 \leftrightarrow \Psi_4$. 
So, let us assume from now on that $\left (\vec{\xi},\vec{k}\right) \neq 0$
and $\Psi_0=0$ everywhere on $U_p$. From the definition of $\pi_0$
it follows that we can choose the
null tetrad globally and uniquely on $U_p$ such that 
$\pi_0$ is purely imaginary (i.e.
$\ov{\pi_0} = - \pi_0$) and that $\left (\vec{\xi}, \vec{k} \, \right) = 1$.
Our next aim is to write down several consequences of the NP equations
in this tetrad on $U_p$. First,
expression (\ref{exP}) can be rewritten on $U_p$ as
\begin{eqnarray}
D  P  = 1, \hspace{1cm}
\Delta  P = - \left (\vec{\xi} , \vec{l} \, \right ) , \hspace{1cm}
\delta P  = \frac{\pi_0}{\ov{P}}, \hspace{1cm}
\ov{\delta}   P  = \frac{\pi_0}{P}.
\label{delP}
\end{eqnarray}
Equations (\ref{eq728}) and (\ref{eq730}) read,
respectively $\epsilon + \ov{\epsilon} = 0$ and
$\ov{\epsilon} \pi_0 + \vec{k} \left ( \pi_0 \right ) - \epsilon \pi_0 = 0$
which, after using $\ov{\pi_0}=- \pi_0$, imply $\epsilon=0$,
$\vec{k} \left (\pi_0 \right )=0$. Let us now define a complex function
 $\alpha_0$ by
\begin{eqnarray*}
\left . \alpha_0 \equiv - \left (\alpha P + \frac{\ov{\pi_0}}{P} \right )
\right |_{U_p}.
\end{eqnarray*}
Applying the commutator (\ref{commdelD}) to $P$ we obtain $\beta
= \ov{ (\alpha_0/P) } $, which introduced in (\ref{eq731}) gives
simply $\vec{k}
\left (\alpha_0 \right ) = 0$. Equation (\ref{expsi4}) reads now
$\Psi_4 = b^2/2 ( P^{-3} \ov{\delta} \pi_0 - 2 P^{-4} \alpha_0 \pi_0  )
|_{U_p}$
and therefore provides an explicit expression for $\Psi_4$.
Similarly, $\left (\vec{\xi} , \vec{l} \, \right ) $ is obtained
from (\ref{rel}) as
$\left ( \vec{\xi} , \vec{l} \right ) = c/2 
- b^2/(4 P) - \ov{b}^2/(4 \ov{P}) - \pi_0^2/(P \ov{P})|_{U_p}$.
Inserting these two expressions into (\ref{eq735}) the
following partial differential equation for $\pi_0$ is obtained
\begin{eqnarray}
\delta \pi_0 = \left .\frac{2 c \left ( \ov{P} - P \right )
+ b^2 - \ov{b}^2 - 8 \pi_0 \ov{\alpha_0}}{4 \ov{P}} \right |_{U_p}.
\label{deltapi0}
\end{eqnarray}
The only NP spin coefficients that remain to be determined are $\gamma$
and $\alpha^0$ (besides $\pi^0$). Regarding $\gamma$, we consider
equation (\ref{eq736}) together with the two equations
obtained by applying the commutators (\ref{commDelD}) and
(\ref{commdelDel}) on $P$. A straightforward combination of these
equations gives
\begin{eqnarray*}
\gamma = \left .\frac{b^2}{4 P^2} + \frac{\pi_0^2}{P^2 \ov{P}}
- \frac{\pi_0 \left ( \alpha_0 + \ov{\alpha_0} \right ) }{ P \ov{P} }
\right |_{U_p} ,
\hspace{1cm}
\vec{\xi} \left ( \pi_0 \right ) |_{U_p} = 0.
\end{eqnarray*}
Regarding $\alpha^0$, equations (\ref{eq740}) and (\ref{eq737}) 
give the following partial differential equations
\begin{eqnarray}
\delta \alpha_0 = \left .
\frac{\left (b^2 - \ov{b}^2 - 2 c P + 2 c \ov{P}
\right ) \left (\ov{\alpha_0} - \alpha_0 \right )
- \pi_0 \left (c + 8 \alpha_0 \ov{\alpha_0} \right )}{4 \pi_0 \ov{P}}
\right |_{U_p} ,
\hspace{1cm} \left .\vec{\xi} \left (\alpha_0 \right ) = 0 \right |_{U_p}.
\label{deltaalpha}
\end{eqnarray}
It is straightforward, although lengthy, to check that the full set
on NP equations on $U_p$ are satisfied provided
$\vec{k} \left (\pi_0 \right ) =0$,
$\vec{k} \left (\alpha_0 \right ) = 0$, 
$\vec{\xi} \left (\pi_0 \right ) =0$,
(\ref{deltapi0}) and
(\ref{deltaalpha}) are fulfilled. So, we must try and solve 
these non-linear partial differential
equations. In order to do that, let us define
a complex vector field $\vec{u} \equiv \ov{P} \vec{m}$.
From (\ref{delP}) and the decomposition (\ref{defyZ}) into real
and imaginary parts we 
have $\vec{k} \left (y \right )=1,
\vec{\xi} \left ( y \right )= 0$, $\vec{u} \left (y \right )= 0$,
$\vec{k} \left (Z \right )=0, \vec{\xi} \left ( Z \right )= 0,
\vec{u} \left ( Z \right ) = - i \pi_0$. 
From $\left (\vec{\xi} , \vec{k} \right ) |_{U_p} =1$ it follows
that $\left \{ \vec{k}, \vec{\xi}, \vec{u}, \vec{\ov{u}} \right \}$
is a basis of vectors at every point in $U_p$. The commutators
between these vectors can be easily computed from 
(\ref{commDelD})-(\ref{commdeldel}) after using the expressions
above. The result reads simply
\begin{eqnarray}
\left .\left [ \vec{k}, \vec{\xi} \right ] = 
\left [ \vec{k}, \vec{u} \right ]  =
\left [ \vec{\xi}, \vec{u} \right ] = \vec{0} \, \, \right |_{U_p},
\hspace{1cm}
\left .\left [ \vec{u}, \vec{\ov{u}} \right ] = - 2 \mbox{i} Z \vec{\xi}
+ 2 \alpha_0 \vec{u} - 2 \ov {\alpha_0} \, \vec{\ov{u}} \, \,\right |_{U_p}.
\label{comuub}
\end{eqnarray}
Now, $dy |_{U_p}$ is nowhere zero because
$\vec{k} \left ( y \right ) =1$ and therefore
$U_p$ can be foliated by the family of hypersurfaces $\{ \Sigma_{y_0} \}$
defined by $y = y_0 = \mbox{const}$. Obviously, the three vector fields 
$\vec{\xi}, \vec{u}, \vec{\ov{u}} $ are tangent to each 
hypersurface $\Sigma_{y_0}$.
Since $\vec{\xi}$ and $\vec{k}$ commute, these two vector fields
are surface-forming. Moreover $\vec{\xi}$ and $\vec{k}$ have
non-zero scalar product on $U_p$ and hence the vector space they generate
at each point $q \in U_p$ is two-dimensional. We can consider the
quotient space of $U_p$ with respect to the two-surfaces generated
by $\vec{\xi}$ and $\vec{k}$. This quotient will be
denoted by 
$S_p = U_p / (\mbox{span}\left \{\vec{\xi}, \vec{k} \right \})$.
By choosing $U_p$ sufficiently small we can assume
that $S_p$ is a differentiable manifold.
The triple $\left ( U_p, S_p, \pi_p \right )$, where
$\pi_p : U_p \rightarrow S_p$ is the canonical projection, is a
bundle over $S_p$. After restricting $U_p$ further if necessary, we
can assume that this bundle is a trivial fiber bundle, i.e. 
$U_p$ is diffeomorphic to
$I_p \times \hat{I}_p \times S_p $, where $I_p$, $\hat{I}_p$ are
open intervals of the real line. 
Any vector field $\vec{v}$ on $U_p$ satisfying $\left [\vec{\xi},
\vec{v} \right ]= \left [ \vec{k}, \vec{v} \right ] = 0$
defines a vector field $\pi_{\star} \vec{v}$
on $S_p$ by projection. Similarly, 
scalar functions on $U_p$ which are constant along both $\vec{\xi}$
and $\vec{k}$ define scalar functions on $S_p$. In order to simplify
the notation we will keep the same symbol when a function is projected
onto $S_p$. The meaning of the expression should be clear from the context.

As $\left [ \vec{u}, \vec{\xi} \, \right ] = 
\left [ \vec{u}, \vec{k} \, \right ] = 0$, we can define
 $\pi_{\star} \vec{u}$. Similarly, $\alpha_0$,
$\pi_0$ and $Z$ can also be projected onto $S_p$. Let us take an
arbitrary coordinate system $\left \{ x^1, x^2 \right \}$ on $S_p$  
and define the complex vector field $\vec{s}$ on $S_p$ by
$\vec{s} = \partial_{x^1} + i \partial_{x^2}$. Since $dZ$ is nowhere
zero on $X$ (because $\vec{u} (Z) = - i \pi_0$) it follows that
$\vec{s}(Z) \neq 0$ everywhere on 
$S_p$. It is convenient to define a real function $N$ on $S_p$ by 
$\pi_0 = (i/2) N \sqrt{\vec{s} (Z) \vec{\ov{s}} (Z)} $ where the square root 
is defined with positive sign. Using $\vec{u} (Z) = - i \pi_0$
 we immediately have 
\begin{eqnarray}
\left .\pi_{\star} \vec{u}   =
\frac{N \sqrt{ \vec{s} \left (Z \right ) \vec{\ov{s}} \left ( Z
\right )}}{2 \vec{s} \left (Z \right ) } \vec{s} \, \, \right |_{S_p}.
\label{uproj}
\end{eqnarray}
Projecting equation (\ref{deltapi0})
onto $S_p$ we can obtain an expression for $\alpha_0$ in terms
of $N$, $Z$ and their derivatives. It reads
\begin{eqnarray}
\alpha_0 = \left ( - \frac{1}{4} \frac{\vec{\ov{s}} (N)}{\vec{\ov{s}} (Z)}
 \sqrt{ \vec{s} (Z) \vec{\ov{s}} (Z) } -
\frac{N}{8 \sqrt{\vec{s}(Z) \vec{\ov{s}}
 (Z)}} \left [  \frac{}{} \vec{s}( \vec{\ov{s}}(Z)) +
\hspace{2cm} \right . \right . \nonumber \\
\left . \left . \left . \hspace{1cm} 
\frac{}{} + \frac{\vec{s}(Z)}{\vec{\ov{s}} (Z)} \vec{\ov{s}} (\vec{\ov{s}} (Z))
+ \frac{8 c Z}{N^2} + \frac{2 i \left (b^2 - \ov{b}^2 \right )}{N^2}
\right ] \right ) \right |_{S_p}.
\label{alpha0}
\end{eqnarray}
The projection  of $\left [ \vec{u}, \vec{\ov{u}} \right ]$
on $S_p$ is 
(see Eq. (\ref{comuub}))
$\left [ \pi_{\star} \vec{u}, \pi_{\star} \vec{\ov{u}} \right ] = 2 \alpha_0 
(\pi_{\star} \vec{u})  - 2 \ov{\alpha_0} (\pi_{\star} \vec{\ov{u}}) |_{S_p}$. 
Inserting (\ref{uproj}) and (\ref{alpha0}) into this commutator and
using $\left [ \vec{s}, \vec{\ov{s}} \right ] = 0$,  
we obtain after a rather long calculation, the remarkably simple equation
\begin{eqnarray}
\vec{s} \left ( \vec{\ov{s}} \left ( Z \right ) \right ) = - \frac{4}{N^2}
\left [ c Z + \frac{i}{4} \left (b^2 - \ov{b}^2 \right ) \right ],
\label{eqZ}
\end{eqnarray}
which is a second order, {\it linear} partial differential equation for $Z$.
We still need to find an equation for $N$. This is accomplished by projecting
the equation for $\delta \alpha_0$ (see  Eq. (\ref{deltaalpha}))
onto $S_p$ and using the expressions for $\alpha_0$ and $\pi_0$ above.
Now the calculation is quite long, but the result is surprisingly simple,
namely
\begin{eqnarray}
N \vec{s} \left ( \vec{\ov{s}} \left ( N \right ) \right ) - \vec{s}
 \left ( N \right )
\vec{\ov{s}} \left ( N \right ) = 2 c.
\label{eqN}
\end{eqnarray}
This is a non-linear partial differential equation which involves
the function $N$ only. Moreover, this equation can be explicitly solved
as follows. Let us introduce the following
metric on $S_p$ 
\begin{eqnarray}
dl^2 = \gamma_{\hat{A} \hat{B}} dx^{\hat{A}} dx^{\hat{B}} =
\frac{2}{N^2} \left [
 (dx^1)^2 + (dx^2)^2 \right ], \hspace{1cm} \hat{A},\hat{B}= 1,2,
\label{2dimmetric}
\end{eqnarray}
which has the Ricci scalar 
$R (\gamma) = N \left ( N_{,x^1 x^1}
+  N_{,x^2 x^2} \right ) -  \left ( N_{,x^1} \right )^2
- \left ( N_{,x^2} \right )^2$. Recalling that $\vec{s} \, |_{S_p} =
\partial_{x^1} + i \partial_{x^2}$ we have that equation (\ref{eqN})
becomes $R (\gamma) = 2 c$.
Hence, the two-dimensional space $\left (S_p, \gamma \right)$
is a space of constant curvature equal to $2c$. The local form
of the metric for such space is well-known and therefore 
$N$ can be determined explicitly. The still arbitrary
coordinate system  $\left \{x^1, x^2 \right \}$ on $S_p$ 
can be chosen so that $N$ takes the standard form
\begin{eqnarray}
N = 1 + \frac{c}{2} \left [ (x^1)^2 + (x^2)^2 \right ].
\label{expN}
\end{eqnarray}
In terms of the metric $\gamma$, it is straightforward to check that the linear
PDE for $Z$ (\ref{eqZ}) can be rewritten as equation (\ref{delZ}) in the
proposition.

The next task is to go from the quotient space $S_p$ back to $U_p$.
To do that recall that 
the fibre bundle $\left (U_p, S_p, \pi_p \right )$ is trivial and 
therefore admits a global cross section $\hat{\sigma}$ (i.e. $\hat{\sigma} :
S_p \rightarrow U_p$ such that $\pi_p \circ \hat{\sigma}= Id |_{S_p}$).
Let us construct a coordinate
system on $U_p$ as follows. Take an arbitrary point $q \in U_p$
and consider the hypersurface $\Sigma_{y_q}$ where $y_q \equiv y(q)$. 
Consider the curve tangent to the vector field $\vec{k}$ passing
through the point  $\hat{\sigma} \circ \pi_p (q)$. Since $\vec{k}$ is
transversal to $\Sigma_{y_q}$ and the fiber bundle is trivial it
follows that this curve must intersect $\Sigma_{y_q}$ at a single point
$r$. Now, $r$ and $q$ belong to the same hypersurface $\Sigma_{y_q}$
and they project onto the same point on $S_p$. Since $\vec{\xi}$ is
tangent to $\Sigma_{y_q}$, it follows that there exists an integral line
of $\vec{\xi}$ that connects $q$ with $r$. Choose the parametrization
$\hat{u}$ associated to the tangent vector $\vec{\xi}$ on this
curve  (i.e. $\vec{\xi} = \partial_{\hat{u}}$ on this curve)
and such that $\hat{u}=0$ on $r$. We define  
a function $u(q)$ as the value of $\hat{u}$ at $q$.
Finally, let $x^1(q)$ and $x^2(q)$ be the values of the coordinates
$\left \{x^1, x^2 \right \}$ of the point $\pi_p (q)$ on $S_p$.
It is easy to check that $\left \{ u(q), y(q), x^1(q), x^2(q) \right \}$
thus constructed is a well-defined coordinate system on $U_p$ and
that, in this coordinate system,  $\vec{\xi} |_{U_p} = \partial_u$ and
$\vec{k} |_{U_p} = \partial_y$. Regarding the vector field $\vec{u}$,
we know that $\vec{u} (y) = 0$ and that its projection onto $S_p$ is
given by (\ref{uproj}). It follows that
\begin{eqnarray}
\left . \vec{u} \right |_{U_p} = 
\frac{N \sqrt{\vec{s}(Z) \vec{\ov{s}} (Z)}}{2 \vec{s}(Z)} \left (
\partial_{x^1} + i \partial_{x^2} \right ) +  \frac{
N \left (B - i A \right ) \sqrt{\vec{s}(Z) \vec{\ov{s}} (Z)}}{2} \partial_u,
\label{vecu}
\end{eqnarray}
where $A, B$ are real functions on $U_p$ which remain to be determined.
From $\left [\vec{u}, \vec{\xi} \, \right ] = \left [\vec{u},
\vec{k} \, \right ] = 0$ it follows $\vec{\xi} (A)=\vec{\xi} (B)
=\vec{k} (A)=\vec{k} (B)= 0 $ and
hence $A$, $B$ define functions on $S_p$. 
The last commutator in (\ref{comuub}) on $U_p$ can be rewritten after
a straightforward calculation as
\begin{eqnarray*}
\left < dZ, dA \right > - \left <  \star_{\gamma} dZ , d B \right > + 2 Z
- A \left ( 2 c Z - 2 b_1 b_2  \right ) =0,
\end{eqnarray*}
where $<,>$ denotes scalar product with respect to the metric
(\ref{2dimmetric}) and $\star_{\gamma}$ is the Hodge dual
with respect to the same metric (recall that $Z$, $b_1$ and $b_2$ are
defined by $P= y + i Z$ and $b= b_1 + i b_2$). This equation
is exactly the linear partial differential equation for $A$ and $B$
appearing in the proposition.
So,  we have constructed
a coordinate system on $U_p$ and have obtained the explicit form of
$\vec{\xi}, \vec{k}, \vec{u}$ and $\vec{\ov{u}}$ in this
coordinate system. Moreover, the scalar products of these vectors
can be determined directly from their definition in terms of the null tetrad
$\left \{\vec{k},\vec{l}, \vec{m}, \vec{\ov{m}} \right \}$. Thus,
the spacetime metric on $U_p$ can be explicitly written. After some algebra,
it is not difficult  to check that this metric
can be written in 
the compact form given in (\ref{metric}). This proves the local isometry 
claimed in the proposition. $\hfill \Box$.

\vspace{5mm}

The explicit expression for $\Psi_4$ for the
metric (\ref{metric}) in the open set $\Mpi \cup \Mk$
will be needed later. Using the expressions obtained in the
proof of Proposition \ref{funda}  it is straightforward to check that
the following expression holds on $\Mpi \cup \Mk$
\begin{eqnarray}
\Psi_4 = \frac{ i \left ( b_1 + i b_2 \right )^2}{ 2
 \left ( y + i Z \right )^4 } \left [ \left < \frac{ dG}{2 G}, dZ +
i \star_{\gamma} dZ \right > + c Z - b_1 b_2 \right ],
\label{psi4}
\end{eqnarray}
where $G \equiv \left < dZ, dZ \right >$. 
The  metric (\ref{metric}) contains several parameters and one
function $Z$ defined on the two-surface $S$. The partial differential
equation for $Z$ is of elliptic type but $S$ may be non-compact
(for instance, $S$ could the hyperbolic
plane). So, in general,
this metric contains arbitrary functions (which can be thought, for instance,
as the boundary conditions of the partial differential equation
(\ref{delZ})).
Thus, we are far from having proven the local isometry
with the Kerr metric which is claimed in Theorem \ref{Main}. 
The reason is, of course, that we have made no use of asymptotic
flatness in Proposition \ref{funda}. This fact shows, in particular,
that asymptotic flatness is an essential ingredient of Theorem \ref{Main}. 
The next lemma exploits the asymptotic conditions
in order to show that $\MD$ is either empty or that it
covers the whole of $\M_3$. 

\begin{lemma}
Let $\M_3$ and $\MD$ be defined as in (\ref{defset}). Then either
$\MD = \emptyset$ or $\MD = \M_3$.
\label{todoonada}
\end{lemma}
{\it Proof:} Since $\MD$ is open, the conclusion of the lemma
is equivalent to $\partial \MD = \emptyset$ in the
topology of $\M_3$. Thus, let us assume
that $\partial \MD$ is non-empty and let us find  a contradiction.
From the fact that
$\Psi_0 = \Psi_4 = 0$ on $\MD$ we find that the Weyl tensor
takes the form
\begin{eqnarray}
\left . \C_{\alpha\beta\gamma\delta} \right |_{\MD} =
\left . 
3 \Psi_2 \left ( \bm{W}_{\alpha\beta} \bm{W}_{\gamma\delta}
+ \frac{4}{3} \I^{\alpha\beta}_{\,\,\,\,\,\,\,\,\gamma\delta} \right )
\right |_{\MD}, \label{DecomD}
\end{eqnarray}
where we used (\ref{Decom}) and the identity
$4 \I^{\alpha\beta}_{\,\,\,\,\,\,\,\,\gamma\delta} = -  \bm{W}^{\alpha\beta}
\bm{W}_{\gamma\delta} + 2 \left ( \bm{U}^{\alpha\beta} \bm{V}_{\gamma
\delta} + \bm{V}^{\alpha\beta} \bm{U}_{\gamma\delta} \right )$ (recall
that $\I^{\alpha\beta}_{\,\,\,\,\,\,\,\,\gamma\delta}$ is the
identity on the space of self-dual two-forms). Thus,
the Weyl tensor on $\MD$ satisfies the hypotheses of Theorem \ref{Old}.
Furthermore, $\M_3 (= \Mun)$ contains the asymptotically
flat end $\Minf$. From the asymptotic conditions it follows that
$\Fsq$ tends to zero at infinity. Using 
(\ref{P}) we find that $1/P$ also tends to zero at infinity (recall that 
$\Fsq = - 4 f^2$). Consequently, equation
(\ref{chi}) implies that $\chi$ tends to $c$ at infinity.
On the other hand, recall that the Killing vector $\vec{\xi}$ was normalized
so that its norm $\lambda \rightarrow 1$ when we approach infinity
in $\Minf$. This implies $c=1$. Moreover, a simple consequence of (\ref{P})
and (\ref{chi})  is
$\Fsq = - b^{-4}(1 -\chi)^4$. The asymptotic behaviour of $\Fsq$ near
infinity forces $b^4=  4 M^2 > 0$ where $M$ is the Komar mass of
$\vec{\xi}$ in $\Minf$ (which is non-zero by hypothesis 2 of 
theorem \ref{Main}). Thus, Theorem \ref{Old}
can be applied to conclude that $\MD$ is locally isometric to the Kerr
spacetime. 

On the other hand, the Kerr
spacetime obviously satisfies all the hypotheses of
Theorem \ref{Main}, and therefore the objects 
$\vec{k}$, $\vec{l}$, $y$, $Z$, $\dots$ that we
constructed on $\M_3$ can also be defined on the Kerr spacetime $\M_{M,a}$.
The subscript $Kerr$ will be used to denote them.
It is a matter of simple calculation using the Kerr geometry
to find that
$y_{Kerr}$ and $Z_{Kerr}$ are
$y_{Kerr} = r$ and $Z_{Kerr} = a \cos \theta$ where
$a$ is the specific angular momentum and $r$ and $\theta$
are standard Boyer-Lindquist coordinates of the Kerr spacetime. 

From our assumption that $\partial \MD \neq \emptyset$ and the
fact that $\MPq \cup \MPz \cup \MD$ is dense in $\M_3$, it follows
that either  $\MPq$ or $\MPz$ is non-empty.
The case $\MPz \neq \emptyset$ can be treated
similarly to $\MPq \neq\emptyset$ just
by interchanging $\vec{k} \leftrightarrow \vec{l}$,
$\vec{m} \leftrightarrow \vec{\ov{m}}$ and $\Psi_0 \leftrightarrow \Psi_4$.
So, we can assume without loss of generality that $\MPq$ is
non-empty and therefore 
$\partial \MD \cap \partial \MPq \neq \emptyset$. 
Take a point $q \in \partial \MD \cap \partial \MPq$ and a sufficiently small
neighbourhood $U_q$ of $q$ such that the null tetrad
$\left \{ \vec{k}, \vec{l}, \vec{m}, \vec{\ov{m}} \right \}$ is smooth
on $U_q$ (hence $\Psi_4$ is also smooth on $U_q$). Define 
$\Xi_q \equiv U_q \cap \partial \MD \cap \partial \MPq$. 
Combining the general
Bianchi equation (\ref{Bian763}) with the expressions for the NP coefficients
(\ref{NPcoeff}) we easily obtain
\begin{eqnarray}
\left . D \Psi_4 = - 4 \Psi_4 \left ( \epsilon + \left ( \vec{\xi},
\vec{k} \, \right )/P \right ) \right |_{U_q}.
\label{dp4}
\end{eqnarray}
Since $\epsilon + \left ( \vec{\xi},\vec{k} \, \right )/P$ is
smooth on $U_q$, we can conclude that an integral line of $\vec{k}$
is either contained  in $\Xi_q$ or else does not intersect $\Xi_q$
(this follows from integrating equation (\ref{dp4}) along any
integral curve of $\vec{k}$, which shows that $\Psi_4$ cannot become
zero along the curve). Moreover, we know
that the Petrov type of any spacetime cannot change along an integral curve
of a Killing vector (see e.g. \cite{Hall}).
This implies that such a curve is also either
contained in $\Xi_q$ or else does not intersect $\Xi_q$. In addition, since
$\Xi_q$ is the boundary of an open subset of $U_q$ (namely $U_q \cap
\MD$) it follows that $\Xi_q$ cannot be contained in any two-dimensional
surface. With this information at hand we wish
to show that there exists an
open subset $U \subset U_q$ with the following
properties: $U \cap \MD \neq \emptyset$, $U \cap \MPq \neq \emptyset$ and
such that that $\left (\vec{\xi}, \vec{k} \, \right )$is non-zero
everywhere on $U \cap \MD$ and on $U \cap \MPq$. Indeed,
if there exists a point $s \in \Xi_q$ where 
$\left ( \vec{\xi}, \vec{k} \, \right ) |_s \neq 0$ the claim is trivial
(a suitable neighbourhood of $s$ would do it). So, we only need
to consider the case when 
$\left . \left ( \vec{\xi}, \vec{k} \, \right ) 
\right |_{\Xi_q}=0$. We know that the metric on $\MD$
is locally isometric to the Kerr metric, so let us evaluate
$W_1 \equiv (\vec{\xi},\vec{k}) (\vec{\xi}, \vec{l} ) (y^2 + Z^2 )$
in the region $\MD \cap U_q$ where the metric is known. We obtain
\begin{eqnarray}
\left .W_1  \right |_{\MD \cap U_q} =
 \left . 2 \left (\vec{\xi},\vec{k} \right ) \left (\vec{\xi}, \vec{l} 
\right) (y^2 + Z^2 ) \right |_{\MD \cap U_q}
 = \left .  y^2 - b^2 y + a^2 \right |_{\MD \cap U_q}.
\label{exppp}
\end{eqnarray}
By assumption $W_1$ vanishes at $\Xi_q$. This implies
$b^4 \geq 4a^2$ and $y |_{\Xi_q} = y_{\pm} \equiv 
(1/2) (b^2 \pm \sqrt{b^4 - 4 a^2 })$. So, the set $\Xi_q$ must be 
contained on one of the event horizons of the Kerr
spacetime. Using standard properties of the Kerr metric, it is immediate to
see that the following expression holds
\begin{eqnarray}
\left .\nabla_{\alpha} W_1 \right |_{y_{\pm}} = \left .
 \pm \sqrt{b^4 - 4 a^2} \, \nabla_{\alpha} y \right |_{y_{\pm}}.
\label{nabW1}
\end{eqnarray}
Furthermore $\nabla_{\alpha} y$ vanishes in the Kerr spacetime
at most on  a two-surface (the bifurcate horizon). So we can always
find $s \in \Xi_q$ so that $\nabla_{\alpha} y |_s \neq 0$ (here
we use the fact that $\Xi_q$ cannot be contained in any two-surface). Take
a neighbourhood of $U \subset U_q$ of $s$ where $\nabla_{\alpha} y |_{U}$
is nowhere zero. If $b^4 - 4 a^2 \neq 0$ (i.e. the Kerr metric is non-extreme)
then (\ref{nabW1}) immediately implies that $W_1$ changes sign
across $\Xi_q \cap U \neq 0$. It follows that 
$\left (\vec{\xi}, \vec{k} \, \right )$ must be non-zero on $\MPq \cap
U$ as claimed. On the other hand, if $b^4 = 4 a^2$, the argument
can be repeated using $W_2 \equiv (W_1)^{1/2}$ instead (in this case,
$W_2 |_{U_q \cap \MD} = y^2 - b^2/2$) and the same
conclusion follows. In any case, we have
that $\left (\vec{\xi}, \vec{k} \, \right )$ must be non-zero on  $\MPq \cap
U$ as claimed.

At this point two cases must be distinguished depending on whether
$a=0$ or $a \neq 0$. The case $a=0$ is very simple because then $Z$
vanishes  on $\MD \cap U$ (and therefore the metric on this open set
is locally the Kruskal-Schwarzschild metric). Using the fact that
the vector field $\vec{l}$ is transversal to $\Xi_{q} \cap U$
(this follows from the geometry of the event horizons of Schwarzschild)
and the equation $\vec{l} (Z) = 0$ (which holds everywhere on $\Mun$ as
a trivial consequence of (\ref{exyZ})) we conclude that $Z=0$ everywhere
on $U$.
So, the Killing vector $\vec{\xi}$ is static on $U$ (because $P$ real
implies (\ref{chi}) $\chi$ real). It is well-known that Petrov type II
is incompatible with a static Killing vector. This excludes the case $a=0$.
Regarding $a \neq 0$, we know from the
Kerr geometry that $\vec{k}$ and $\vec{\xi}$ are linearly independent
on the event horizon. Furthermore, from standard
properties of the Kerr geometry
we find that  $\eta_{\alpha\beta\lambda\mu} \xi^{\beta} k^{\lambda}
l^{\mu}$ vanishes in the Kerr spacetime  only at points where
$\vec{\xi} = \vec{0}$ or at the axis of symmetry of the axial Killing,
which is also a two-dimensional surface.
Thus, there exists a point $p \in \Xi_q \cap U$ where
$\vec{\xi}$ and $\vec{k}$ are linearly independent and that 
$\eta_{\alpha\beta\lambda\mu} \xi^{\beta} k^{\lambda} l^{\mu} |_p \neq 0$.
Hence, there exists an open neighbourhood $U_p \subset U $ of $p$
where $\eta_{\alpha\beta\lambda\mu} \xi^{\beta} k^{\lambda} l^{\mu} |_{U_p}
\neq 0$ and such that the quotient
$S_{p}=U_p/\mbox{span} \{\vec{\xi},\vec{k} \}$ is
a differentiable manifold. As before, we will denote by $\pi_p$
the standard projection of $U_p \rightarrow S_{p}$.
As in Proposition \ref{funda} we endow $S_{p}$ with a metric $\gamma$ of
constant curvature equal to 2 (recall that $c=1$ from asymptotic flatness).
On the open region $U_p \cap \MPq$, Proposition \ref{funda} can
be applied and therefore the local metric on
$U_p \cap \MPq$ is given by (\ref{metric}). Since $c=1$ and $b_1 b_2 =0$
(recall that $b^4 = 4 M^2$) we have that $Z$
satisfies $\Delta_{\gamma} Z = -2 Z$ on 
$\pi_p (U_p \cap \MPq)$. On the other hand, the metric
on $\MD \cap U_p$ is locally Kerr. Therefore $Z |_{\MD} = a \cos{\theta}$
which clearly satisfies $\Delta_{\gamma} Z= -2 Z$ on 
$\pi_p (\MD \cap U_p) $. By construction we have
that $\pi_p (\MD \cap U_p)$ and $\pi_p (\MPq \cap U_p)$
are both non-empty and dense in $S_{p}$. Since $Z$ is
smooth on $S_{p}$ it follows that 
$\Delta_{\gamma} Z= -2 Z$ holds everywhere on $S_{q}$. Now,
$\Delta_{\gamma} Z= -2 Z$ is a
second order {\it elliptic} partial  differential equation. Hence,
\cite{Gara} $Z$ is analytic on $S_{p}$. Since $Z = a \cos \theta$ on the
open non-empty subset of $S_p$
we can conclude that  $Z = a \cos \theta$ 
everywhere on $S_{p}$. This gives the contradiction we need
because on the one hand we have
$\Psi_4 \neq 0$ on $\M^4_3 \cap U_p$ and on the other hand
substituting $Z= a \cos \theta$ into (\ref{psi4}) we find
$\Psi_4 |_{U_p} = 0$. Thus we can conclude that  
$\partial \MPq \cap \partial \MD = \emptyset$
(and $\partial \MPz \cap \partial \MD = \emptyset$ by a 
similar argument) and the proof of the lemma is completed. $\hfill \Box$

\vspace{5mm}

So, we already know that $\MD$ is either
empty or covers $\M_3$ and that the metric on $\MD$ is locally
isometric to Kerr. In order to complete the proof of Theorem \ref{Main} we
need to show first that $\MD = \emptyset$ is impossible
and second that $\Mun = \M$. Proving the first claim uses asymptotic
flatness in an essential way.

\vspace{5mm}

{\it \bf Proof of the Theorem}

\vspace{5mm}

We must first exclude the possibility that $\MD = \emptyset$. 
So, we suppose $\MD = \emptyset$ and obtain a contradiction.
Notice first 
that $\MD = \emptyset$ does not imply yet that one of $\Psi_0$ or
$\Psi_4$ is everywhere zero. It is still possible that
$\Psi_0$ and $\Psi_4$ are nonzero on some (necessarily disjoint)
regions of $\M_3$
(i.e. that $\MPq \neq \emptyset$ and $\MPz \neq \emptyset$).
We know that $\M_3 = \Mun$
and therefore there exists an asymptotically flat end $\Minf \subset
\M_3$. By restricting $\Minf$ if necessary we can assume
that $\vec{\xi}$ is timelike in $\Minf$. It is well-known
\cite{MZH} that a vacuum spacetime admitting a Killing vector
field is analytic in the region where the Killing vector
is timelike. Consequently, in the strictly stationary
region (i.e. where $\vec{\xi}$ is timelike)
the vector fields $\vec{k}$ and $\vec{l}$ are analytic. This follows
from the fact that they are the eigenvectors corresponding
to simple eigenvalues of an analytic endomorphism and therefore
no multiplicity changes are permitted). Thus, there exists
(locally) an analytic null tetrad
$\{ \vec{k}, \vec{l}, \vec{m}, \vec{\ov{m}} \}$.
In the local neighbourhood where this tetrad is analytic,
$\Psi_0$ and $\Psi_4$  are also analytic. Thus, we can conclude 
that in the asymptotically flat end $\Minf$ 
(where $\vec{\xi}$ is timelike), either $\Psi_0$ or
$\Psi_4$ vanish everywhere (because at least one of them vanishes
 on an open non-empty set). As before,
the two cases $\Psi_0=0$ or $\Psi_4=0$
are related to each other  by a transformation $\vec{k}
\leftrightarrow \vec{l}$,
$\vec{m} \leftrightarrow \vec{\ov{m}}$, $\Psi_0 \leftrightarrow
\Psi_4$ and hence we can assume without loss of generality that
$\Psi_0$ vanishes everywhere on $\Minf$ (i.e. $\Minf \subset
\MPq$). From the
fact that $\left [ \vec{\xi}, \vec{k} \right ]  \propto
\vec{k}$ (see (\ref{comxik})) and that $\{ \vec{\xi},
\vec{k} \}$ span a two-dimensional vector subspace
at every point in $\Minf$, 
it follows that $\vec{\xi}$ and $\vec{k}$ are tangent to families
of two-surfaces. Furthermore, $\Minf$ can be chosen to have
topology $\Bbb{R}
\times \Bbb{R} \times S^2$ ($S^2$ is the two-sphere).
Using the fact that $\vec{k}$ is
a principal null direction of the Weyl tensor (this follows from
(\ref{Decom}) after taking into account that $\Psi_0 = 0$), asymptotic
flatness implies 
(possibly after restricting $\Minf$) that
$\Minf / \mbox{span} \{ \vec{\xi}, \vec{k} \} \simeq S^2$.
Let us introduce on $S^2$ the round metric $\gamma$ with constant
curvature $R = 2$. So $(S^2, \gamma)$ is a complete
and simply connected Riemannian space. 
Let us also define the canonical projection onto this quotient
by $\tilde{\pi} : \Minf \rightarrow S^2$. 
Proposition \ref{funda} takes care of the local form of the metric
at points where $\eta_{\alpha\beta\lambda\mu} \xi^{\beta} k^{\lambda}
l^{\mu} \neq 0$ (i.e. at points on $\Mpi$). So, we must analyze what
happens on the set of points where $\eta_{\alpha\beta\lambda\mu}
\xi^{\beta} k^{\lambda} l^{\mu} = 0$. To do that, we
define the set $T = \{ p \in \Minf ; \,\,
\eta_{\alpha\beta\lambda\mu} \xi^{\beta} k^{\lambda}l^{\mu} |_p = 0 \, \}$
and prove that $T$ must have empty interior. Indeed,
suppose there exists a point $p \in \stackrel{o}{T}$ and take
a sufficiently small open neighbourhood $U_p \subset \stackrel{o}{T}$
of $p$ where the null
tetrad $\{\vec{k}, \vec{l}, \vec{m}, \vec{\ov{m}} \}$ is well-defined.
We have  (from the definition of $T$) that $(\vec{\xi},\vec{m} ) |_{U_p}=0$.
Since $U_p$ is open, equation (\ref{expsi4})
implies that $\Psi_4 |_{U_p} =0$.
Hence we have $\Psi_0 |_{{\small \stackrel{o}{T}}} = 
\Psi_4 |_{{\small \stackrel{o}{T}}} = 0$ and
therefore $\stackrel{o}{T} \subset 
\MD$ which we are assuming is empty. Therefore $T$
has empty interior. Thus, the set 
$\Minf \cap \Mpi $ is dense in $\Minf$.
Let us prove that its projection is also dense on $S^2$. Indeed, a
trivial consequence of (\ref{comxik}) is
${\pounds}_{\vec{\xi}}
 \, (\eta_{\alpha\beta\lambda\mu} \xi^{\beta} k^{\lambda}l^{\mu})
 |_{\Minf} = 0$.
Similarly, taking into account that, locally,
$\eta_{\alpha\beta\lambda\mu} \xi^{\beta} k^{\lambda} l^{\mu} =
i \left [\left ( \vec{\xi \,} , \vec{\ov{m}} \right ) m_{\alpha} - 
\left ( \vec{\xi \,} , \vec{m} \right ) \ov{m} _{\alpha} \right ]$,
equation (\ref{Dxim}) shows  that the vanishing of
$\eta_{\alpha\beta\lambda\mu} \xi^{\beta} k^{\lambda}l^{\mu}$ at any
point in $\Minf$ implies its vanishing everywhere along the integral lines of 
$\vec{k}$. Putting these two things together, we can conclude
$\tilde{\pi}^{-1} \circ \tilde{\pi} \left( T \right ) = T$, i.e.
the set $T$ is the anti image of certain subset of $S^2$. Therefore
$\tilde{\pi} (\Minf \cap \Mpi)$
is dense in $S^2$. Recalling that $\Minf \subset \Mk$, we observe that
Proposition \ref{funda} can be applied on $\tilde{\pi} (\Minf \cap \Mpi)$.
Taking into account that $c=1$ and $b_1 b_2 = 0$,
we find from Proposition \ref{funda} that $Z$ fulfills
$\Delta_{\gamma} Z = - 2 Z$ on a
dense subset of $S^2$ and hence everywhere. Thus,
 $Z$ satisfies the linear equation $\Delta_{\gamma} Z = - 2 Z$
on the whole two-sphere.
This equation is an eigenvalue equation (with eigenvalue equal to $-2$)
for the Laplacian on the 
round sphere. Since $Z$ is defined in terms of $P$, which is a smooth
function on $\M_3$, the solution of this eigenvalue equation must be regular
and smooth everywhere on  $S^2$. The eigenvalues of the Laplace-Beltrami
operator on the sphere are well-known and the general regular solution of
$\Delta_{\gamma} Z = - 2 Z$ can always be
written as $Z = a \cos \theta$ where $a$ is a real constant
and $\theta$ is a suitably chosen colatitude angle on the sphere.
This expression  for $Z$ shows that $\Psi_4$ (given in (\ref{psi4})) must be
zero everywhere on $\Minf \subset \M_3$. Since $\Psi_0$ vanishes also
on $\Minf$, this contradicts the fact that 
$\MD = \emptyset$. Thus, Lemma \ref{todoonada} shows $\MD = \M_3$. 
Since Proposition \ref{Mun=M3} tells us that $\M_3 = \Mun$ we conclude that
$(\Mun, \gun)$ is locally isometric to the Kerr spacetime.

The final part of the proof consists in showing that 
$\Mun = \M$. Assume  that $\Mun$ is a proper subset of $\M$ and take a point
$q$ on the topological boundary of $\Mun$ as a subset
of $\M$. 
From the continuity of $\Fsq$ and the fact that $\Fsq = - b^{-4} (1 - \chi )^4$
on $\Mun$, we must have
$\lambda |_q =1$, hence $\vec{\xi}$ is timelike in some neighbourhood of
$q$.  Consider a smooth curve $\gamma_p (s)$ defined on some interval
$s_0 < s \leq 1$ such that $\gamma_p (s) \in \Mun$, 
$\forall s \in (s_0,1)$,
$\gamma_p(1)=p$ and such that the tangent vector $\vec{\dot{\gamma}} (s)$ is
orthogonal to $\vec{\xi}$ and of unit length (the existence of such a curve
is easy to establish).
Define the real function $Y(s) = (y \circ \gamma_p) (s)$.
Since $\chi \rightarrow 1$ at $p$ we have from (\ref{chi}) that
$P$ diverges at $p$. Taking into account that $Z$ remains bounded,
we find $Y(s) \rightarrow \infty$ when $s \rightarrow 1$. 
In the Kerr metric we have $\nabla_{\alpha} y \nabla^{\alpha} y =
(y^2 - 2 M y + a^2)/(y^2+ a^2 \cos^2 \theta)$ and therefore
$\nabla_{\alpha} y \nabla^{\alpha} y |_{\gamma_p(s)}
\rightarrow 1$ when $s \rightarrow 1$. Hence we can assume
that $\nabla_{\alpha} y |_{\gamma_p(s)}$ is
spacelike for $s \in (s_0,1)$. Then 
\begin{eqnarray*}
\left (\frac{dY}{ds} (s) \right )^2 = \left (
\nabla_{\alpha} y |_{\gamma_p(s)} \dot{\gamma}^{\alpha}_p(s) \right )^2 
\leq \nabla_{\alpha} y \nabla^{\alpha} y |_{\gamma_p(s)}, 
\end{eqnarray*}
where we have used the fact that $\{ \nabla_{\alpha} y |_{\gamma_p(s)},
\dot{\gamma}_p(s) \}$ define a spacelike two-plane  and we have
applied the Schwarz inequality. Hence $\frac{dY}{ds}$ stays bounded, which
contradicts $Y \rightarrow \infty$ when $s \rightarrow 1$.
This completes the proof of the theorem. $\hfill \Box$.

\section*{Acknowledgements}

I wish to thank Ra\"ul Vera and
Jos\'e M.M. Senovilla for a careful reading of the manuscript and for
valuable comments.
This work has been partially supported by projects 
UPV172.310-G02/99 and 1998SGR 00015.

\end{document}